\UseRawInputEncoding
\documentclass[twocolumn, trackchanges]{aastex63}
\usepackage[utf8]{inputenc}
\usepackage{appendix}
\usepackage[normalem]{ulem}
\usepackage{soul}
\usepackage{cancel}
\usepackage{hyperref}
\usepackage{changepage}
\usepackage{outlines}
\usepackage{multirow}
\usepackage{booktabs} 
\usepackage{threeparttable} 
\usepackage{booktabs}
\usepackage{amsmath}

\usepackage{tablefootnote}

\def\priorAUMiclnA{$\mathcal{LU}(0.15, 1500)$}
\def\priorAUMiclnl{$\mathcal{LU}(10, 1000)$}
\def\priorAUMiclngamma{$\mathcal{LU}(0.1, 10)$}
\def\priorAUMiclnP{$\mathcal{LU}(4.8, 4.9)$}
\def\priorAUMiclns{$\mathcal{LU}(0.15, 1500)$}

\def\postLINAUMiclnA{$15.9^{+1.7}_{-2.7}$}
\def\postLINAUMiclnl{$124^{+20}_{-14}$}
\def\postLINAUMiclngamma{$1.38^{+0.36}_{-0.34}$}
\def\postLINAUMiclnP{$4.8620^{+0.0046}_{-0.0029}$}
\def\postLINAUMiclns{$1.64^{+0.10}_{-0.12}$}

\def\postLINAUMicBmaglnP{$4.8636^{+0.0021}_{-0.0019}$}

\def\postLINAUMicModellnA{$7.81^{+0.96}_{-1.32}$}
\def\postLINAUMicModellnl{$160^{+27}_{-32}$}
\def\postLINAUMicModellngamma{$1.67^{+0.43}_{-0.44}$}
\def\postLINAUMicModellnP{$4.8635^{+0.0024}_{-0.0029}$}

\def\RUNIDAUMIC{19AD97, 19AF26, 19AH28, 19AP42, 19BD97, 20AH93, 20AP42, 20BP42, 21AP42, 21BP42, 22AP42, 22BP45, 23AP45, 23BP45}
\def\RUNIDBARNARD{18BE97, 19AP40, 19BP40, 20AP40, 20BP40, 21AP40, 21BP40, 22AP40, 23AQ57, 23BQ57, 24AQ57}
\def\RUNIDHDxxxxx{18BE99, 19AP40, 20AC01, 21BC16, 23AQ57, 23BQ57}
\def\RUNIDepseri{0104.C-0863(A), 072.C-0488(E), 073.C-0784(B), 074.C-0012(A), 074.C-0012(B), 076.C-0878(A), 077.C-0530(A), 078.C-0833(A), 079.C-0681(A), 192.C-0852(A), 60.A-9036(A)}
\def\RMSresidualharps{0.087}
\def\nstarsspirou{44}
\def\nstarsharps{105}
\def\nstarsoverlap{8}
\def\RMSresidualspirou{0.036}
\def\mederrHDdtemp{0.97}
\def\mederrHDmodel{0.39}
\def\pearsondtempbmag{-0.924}
\def\slopedtempbmagfit{<B> = (-0.0086\pm0.0003) {\rm d}{\it Temp} + (2.733\pm0.004)}

\def\priorBarnardlnA{$\mathcal{LU}(0.015, 150)$}
\def\priorBarnardlnl{$\mathcal{LU}(300, 1000)$}
\def\priorBarnardlngamma{$\mathcal{LU}(0.1, 10)$}
\def\priorBarnardlnP{$\mathcal{LU}(100, 170)$}
\def\priorBarnardlns{$\mathcal{LU}(0.015, 150)$}

\def\postLINBarnardlnA{$3.10^{+0.70}_{-1.11}$}
\def\postLINBarnardlnl{$309.0^{+6.8}_{-15.1}$}
\def\postLINBarnardlngamma{$0.83^{+0.32}_{-0.61}$}
\def\postLINBarnardlnP{$153.3^{+3.1}_{-2.4}$}
\def\postLINBarnardlns{$0.493^{+0.037}_{-0.040}$}
\def\priorepsilonEridanilnA{$\mathcal{LU}(0.008, 80)$}
\def\priorepsilonEridanilnl{$\mathcal{LU}(10, 1000)$}
\def\priorepsilonEridanilngamma{$\mathcal{LU}(0.1, 10)$}
\def\priorepsilonEridanilnP{$\mathcal{LU}(8, 20)$}
\def\priorepsilonEridanilns{$\mathcal{LU}(0.008, 80)$}

\def\postLINepsilonEridanilnA{$1.08^{+0.25}_{-0.44}$}
\def\postLINepsilonEridanilnl{$18.6^{+3.6}_{-3.7}$}
\def\postLINepsilonEridanilngamma{$0.74^{+0.30}_{-0.43}$}
\def\postLINepsilonEridanilnP{$12.27^{+0.31}_{-0.36}$}
\def\postLINepsilonEridanilns{$0.023^{+0.012}_{-0.030}$}
\def\omegaepseri{$53.0^{+8.6}_{-7.6}$$^\circ$}

\definecolor{barbie}{RGB}{240, 33, 138}

\newcommand{\magb}{$<$$B$$>$ }
\newcommand{\epse}{$\epsilon$\,Eridani }

\newcommand\dtemp{d\textit{Temp} }

\DeclareRobustCommand{\okina}{\raisebox{\dimexpr\fontcharht\font`A-\height}{\scalebox{0.8}{`}}}

\shortauthors{Artigau et al.}

\begin{document}

\title{Measuring Sub-Kelvin Variations in Stellar Temperature with High-Resolution Spectroscopy}
\correspondingauthor{\'Etienne Artigau}
\email{etienne.artigau@umontreal.ca}

\author[0000-0003-3506-5667]{\'Etienne Artigau}
\affiliation{Institut Trottier de recherche sur les exoplan\`etes, D\'epartement de Physique, Universit\'e de Montr\'eal, Montr\'eal, Qu\'ebec, Canada}
\affiliation{Observatoire du Mont-M\'egantic, Qu\'ebec, Canada}

\author[0000-0001-9291-5555]{Charles Cadieux}
\affiliation{Institut Trottier de recherche sur les exoplan\`etes, D\'epartement de Physique, Universit\'e de Montr\'eal, Montr\'eal, Qu\'ebec, Canada}

\author[0000-0003-4166-4121]{Neil J. Cook}
\affiliation{Institut Trottier de recherche sur les exoplan\`etes, D\'epartement de Physique, Universit\'e de Montr\'eal, Montr\'eal, Qu\'ebec, Canada}

\author[0000-0001-5485-4675]{Ren\'e Doyon}
\affiliation{Institut Trottier de recherche sur les exoplan\`etes, D\'epartement de Physique, Universit\'e de Montr\'eal, Montr\'eal, Qu\'ebec, Canada}
\affiliation{Observatoire du Mont-M\'egantic, Qu\'ebec, Canada}

\author[0009-0004-2993-7849]{Laurie Dauplaise}
\affiliation{Institut Trottier de recherche sur les exoplan\`etes, D\'epartement de Physique, Universit\'e de Montr\'eal, Montr\'eal, Qu\'ebec, Canada}

\author[0000-0002-0111-1234]{Luc Arnold}
\affiliation{Canada-France-Hawai\okina i Telescope, CNRS, Kamuela, HI 96743, USA}

\author{Maya Cadieux}
\affiliation{Institut Trottier de recherche sur les exoplan\`etes, D\'epartement de Physique, Universit\'e de Montr\'eal, Montr\'eal, Qu\'ebec, Canada}

\author[0000-0001-5541-2887]{Jean-Fran\c{c}ois Donati}
\affiliation{CNRS, OMP, Universit\'e de Toulouse, 14 Avenue Belin, F-31400 Toulouse, France}

\author[0000-0003-4019-0630]{Paul Cristofari}
\affiliation{Center for astrophysics $\vert$ Harvard \& Smithsonian, 60 Garden Street, Cambridge, MA 02138, United States}

\author[0000-0001-5099-7978]{Xavier Delfosse}
\affiliation{Observatoire de Gen\`eve, D\'epartement d’Astronomie, Universit\'e de Gen\`eve, Chemin Pegasi 51b, 1290 Versoix, Switzerland}

\author[0000-0002-1436-7351]{Pascal Fouqu\'e}
\affiliation{CNRS, OMP, Universit\'e de Toulouse, 14 Avenue Belin, F-31400 Toulouse, France}

\author[0000-0002-2842-3924]{Claire Moutou}
\affiliation{CNRS, OMP, Universit\'e de Toulouse, 14 Avenue Belin, F-31400 Toulouse, France}

\author[0009-0005-1139-3502]{Pierre Larue}
\affiliation{Univ. Grenoble Alpes, CNRS, IPAG, F-38000 Grenoble, France}

\author[0000-0002-1199-9759]{Romain Allart}
\affiliation{Institut Trottier de recherche sur les exoplan\`etes, D\'epartement de Physique, Universit\'e de Montr\'eal, Montr\'eal, Qu\'ebec, Canada}

\begin{abstract}

The detection of stellar variability often relies on the measurement of selected activity indicators such as coronal emission lines and non-thermal emissions. On the flip side, the effective stellar temperature is normally seen as one of the key fundamental parameters (with mass and radius) to understanding the basic physical nature of a star and its relation with its environment (e.g., planetary instellation). We present a novel approach for measuring {disk-averaged} temperature variations to sub-Kelvin accuracy inspired by algorithms developed for precision radial velocity. This framework uses the entire content of the spectrum, not just pre-identified lines, and can be applied to existing data obtained with high-resolution spectrographs. We demonstrate the framework by recovering the known rotation periods and temperature modulation of Barnard star and AU Mic in datasets obtained in the infrared with SPIRou at CHFT and at optical wavelengths on $\epsilon$ Eridani with HARPS at ESO 3.6-m telescope. We use observations of the transiting hot Jupiter HD189733\,b, obtained with SPIRou, to show that this method can unveil the minute temperature variation signature expected during the transit event, an effect analogous to the Rossiter-McLaughlin effect but in temperature space. This method is a powerful new tool for characterizing stellar activity, and in particular temperature and magnetic features at the surfaces of cool stars, affecting both precision radial velocity and transit spectroscopic observations. We demonstrate the method in the context of high-resolution spectroscopy but the method could be used at lower resolution.

\end{abstract}

\section{Introduction}

The ability to detect and analyze exoplanets has ushered in an era where characterizing their atmospheres, surfaces, and potential habitability stands at the forefront of astronomical pursuits. The study of exoplanets is performed through several technical means. Radial velocity measurements, involving the detection of the kinematic effect on the host
star by the Keplerian motion of its planets, have been the first successful discovery technique. Although this method is still used to identify a large number of planets, it has been overtaken in the sheer number of discoveries by photometric surveys where one looks for the minute flux change during a transit when a planet passes in front of its parent star. While precision radial velocity (pRV) and photometric surveys have contributed to the discovery of the majority of known exoplanets to date, both methods remain strongly biased toward the discovery of close-in planets ($<$0.1\,AU for transit, $<$5\,AU for pRV). Direct imaging surveys with adaptive optics, as well as very deep imaging in the infrared, complement the current portrait of exoplanets with companions now known at separations up to thousands of astronomical units.

As exoplanet pRV searches gained momentum, they were confronted with a series of challenges, one of the most significant being the intricate interplay between the activity of host stars and the Keplerian motion of exoplanets. This complex relationship introduces variability and noise at all timescales, ranging from minutes for oscillation \citep{brewer_bayesian_2007, bazot_estimating_2012} and granulation \citep{kallinger_connection_2014, cegla_stellar_2018}  to years for magnetic cycles \citep{baliunas_chromospheric_1995, gomes_da_silva_long-term_2012}. 
Activity filtering techniques have emerged as crucial tools to mitigate the confounding effects of stellar activity on exoplanet observations, ensuring that signals emerging from stellar variability are not falsely interpreted as kinematic signatures of exoplanets (e.g., Alpha Cen B\,b,{\cite{rajpaul_ghost_2016}; Gl699\,b, \cite{lubin_stellar_2021}; AD\,Leo\,b, \cite{carmona_near-ir_2023}; LHS 1140\,d, \cite{cadieux_new_2024}; TW Hya\,b, \cite{huelamo_tw_2008})}.

High-accuracy photometry, ideally obtained contemporaneously with RV measurements, informs on the distribution of features on the photosphere that depart from the mean brightness of the star. A passing stellar spot or a plage on a rotating star will lead to an RV jitter as it affects the mean line shape while also imparting a photometric signature. One can show that at first order, for a flux $f$ and its time derivative $f'$, the RV signature of the spot and plages will scale as $f \cdot f'$ \citep{aigrain_simple_2012}. An additional term is added to account for the convective blueshift inhibition \citep{meunier_using_2010, meunier_reconstructing_2010}, leading to a correspondence between the flux and its time-derivative and RV jitter. Convective blueshift inhibition refers to the reduction of the net blueshift effect caused by convective motions in a star's photosphere, often due to factors like magnetic activity. This inhibition can introduce noise and systematic errors into RV measurements and becomes dominant against rotation-induced jitter for slowly rotating stars.

Several spectroscopic activity indicators have been used, each tracing a different facet of the activity and, ultimately, statistically tied to the perturbing RV jitter. 

The indicators inform on the underlying physical common cause between an ancillary observable and the line barycenter shifts that mimic Keplerian reflex motions.
Tracers of line profile distortions were first measured from the cross-correlated function (CCF) representing the averages of many spectral lines. These indicators include the bisector inverse slope (BIS; {\cite{queloz_no_2001}}) measuring the skewness/asymmetry of the CCF (see also $V_{\rm span}$; {\cite{boisse_disentangling_2011}}) or the full-width at half maximum (FWHM; {\cite{queloz_corot-7_2009}}) tracing the CCF width change.
Inferred as a differential measurement from a stellar template spectrum, the chromatic index (CRX) and the differential line width (dLW) indicators also reflect variation in line shape from stellar activity \citep{zechmeister_spectrum_2018,artigau_line-by-line_2022}.

{\cite{haywood_unsigned_2022}} estimated the Sun's disk-averaged RV variations using the Fe I line observed by SDO/HMI and modelled the magnetic activity, showing that a linear fit to the unsigned magnetic flux reduces RV variation by 62\%. The findings suggest that unsigned magnetic flux is an excellent proxy for activity-induced RV variations. Other tracers are based on emission in the core of specific lines, such as the $S$-index and $\log R^{\prime}_{\rm HK}$ indicators for the CaII H and K lines on Sun-like stars \citep{noyes_rotation_1984}, or H$\alpha$ line for cooler stars \citep{kurster_low-level_2003}. These chromospheric indicators may, however, only weakly correlate with the RV jitter that arises from photospheric activity \citep{schofer_carmenes_2019}.

We propose a novel technique to measure stellar activity, namely a precise differential temperature measurement based on a high-resolution spectrum, measuring slight variations in the disk-averaged temperature due to the passage of active regions, either hotter than the photosphere (i.e., plages) and/or cooler (i.e., stellar spots). This technique has the advantage of applying to all existing pRV archival data. While the technique was devised with exoplanet searches in mind, it can be applied to datasets obtained with high-resolution spectrographs that are not stabilized at the m/s level.

The paper is structured as follows. In section~\ref{sect:method}, we present the mathematical framework of the method, we then describe the datasets used to build the spectral templates necessary for this method in section~\ref{sect:templates}. In section~\ref{sect:prv}, we apply this framework to datasets obtained with an infrared spectrograph \citep[SPIRou;][]{donati_spirou_2020} and an optical spectrograph \citep[HARPS;][]{pepe_harps:_2002, mayor_setting_2003} in the context of pRV monitoring and transit spectroscopy. We explore the change in the disk-averaged temperature during a transit in Section~\ref{sec:titc}. The proposed method being based on observations, we explore the caveats of using stellar models in our analysis in Section~\ref{sect:models_vs_templates}. We summarize our findings and highlight exciting future research avenues in Section~\ref{sect:conclusions}.

\section{Method\label{sect:method}}

Our approach to temperature measurement rests on the line-by-line technique described in {\cite{artigau_line-by-line_2022}} (LBL) and expands its usage to the temperature domain rather than velocity \citep{al_moulla_measuring_2022}. The LBL technique involves measuring velocity, or equivalently small wavelength shifts, from the first derivative expansion of a nearly noise-free template of a star matched to a given observed spectrum. Crucially, this technique operates independently over individual spectral lines, allowing for the identification and removal of spectral outliers that would otherwise bias the velocity measurement. This technique was initially designed with stellar activity filtering in mind \citep{dumusque_measuring_2018, cretignier_measuring_2020} but also proved to be efficient at removing instrumental signatures \citep{cretignier_yarara_2021, ould-elhkim_wapiti_2023}. While a powerful tool at optical wavelengths, the LBL technique presents an even greater interest in the near-infrared where deep telluric absorption affects RV measurements \citep{figueira_radial_2016, wang_characterizing_2022} and detector cosmetics are notably worse than for CCDs \citep{artigau_h4rg_2018}. Near-infrared LBL velocity measurements reach the $<2$\,m/s accuracy \citep{artigau_line-by-line_2022, artigau_nirps_2024} and are, to date, the only method consistently providing constraints on planetary signatures at near-infrared wavelengths at the m/s-level and better \citep[e.g.,][]{cadieux_toi-1452_2022, gan_tess_2022, moutou_characterizing_2023}.

We introduce an adaptation of the LBL framework by substituting the velocity derivative with the temperature derivative of the spectrum. This modification enables a differential evaluation of temperature relative to a template. Leveraging the full spectrum's information content, this adapted approach prioritizes precision while minimizing error.

Central to the LBL algorithm is a formalism introduced by {\cite{bouchy_fundamental_2001}}, in which one finds the shift $\delta\lambda(i)$ in wavelength space (and correspondingly, in velocity space) between a spectrum $A(i)$ and its (ideally) noiseless template $A_0(i)$ for pixel $i$ as :
\begin{equation}
A(i) - A_0(i) = \frac{\partial A_0(i)}{\partial \lambda(i)} \delta \lambda(i)
\label{eq:b01}
\end{equation}

 Where the  $\frac{\partial A_0(i)}{\partial \lambda(i)}$ term corresponds to the velocity gradient of the template (i.e., change in flux for an infinitesimal velocity offset). This can be seen as the first term of a Taylor expansion of flux against wavelength. This can be expressed as a velocity gradient by using

 \begin{equation}
 \frac{\delta V}{c} = \frac{\delta\lambda}{\lambda}
 \end{equation}
 
 Where $\delta V$ is the velocity shift, $\delta\lambda$ the wavelength shift and $c$ the speed of light. Within this framework, one can take the derivative of the spectrum against any relevant quantity. Here we explore the derivative in temperature,  $\frac{\partial A_0(i)}{\partial T(i)}$. While the derivative against velocity can be determined from a high signal-to-noise template, the derivative in temperature requires a library of spectra of objects with known temperatures. By analogy to the velocity-domain weighting of residuals presented in {\cite{bouchy_fundamental_2001}}, one can derive a per-pixel temperature change expressed as:
\begin{equation}
\delta T(i) =  \frac{A(i) - A_0(i)}{\partial A_0(i) /\partial T(i) },
\label{eq2}
\end{equation}

and a corresponding error in the temperature as derived for that spectral element;

\begin{equation}
\sigma_T(i) =  \frac{\sigma(i)}{\partial A_0(i) / \partial T(i) }
\label{eq:noise}
\end{equation}
Here, $\sigma(i)$ is defined as the running standard deviation of the difference between the spectrum of interest and the corresponding noiseless template.

One must note that the velocity and temperature derivatives may have a cross-term: if the velocity of the template and that of the spectrum of interest differ, then there will be an error in the determination of $\delta T$. We, therefore, compute the $\delta T$ after convergence of the velocity iterations of the LBL. While a small ($\ll 1$ line width) velocity offset does not impact temperature measurements, larger velocity residuals will lead to a mismatch between the temperature gradient spectrum and the spectrum being analyzed.

Considering that an optimal weighted sum is performed with weights proportional to the inverse of the square of uncertainties, one therefore has a mean temperature change for the domain of interest. {This change in the disk-averaged temperature is designated as \dtemp and determined in a weighted sum through the following relation:

\begin{equation}
{\rm d}{\it Temp} = \frac{\sum  \delta T(i)  \sigma_T(i)^{-2} }{\sum \sigma_T(i)^{-2}}
\label{sum_rv}
\end{equation}
}
The  uncertainty of a weighted sum becomes:
{
\begin{equation}
\sigma_{{\rm d}{\it Temp}}= \frac{1}{\sqrt{\sum \sigma_T(i)^{-2}}}
\label{err_rv}
\end{equation}
}

We keep the framework of the LBL and opt for a simple mixture model (see Appendix~B in \cite{artigau_line-by-line_2022}) to compile all per-line measurements into a final temperature estimate.

One key element in the above framework is to determine the derivative as a function of temperature, which requires a library of relatively high-SNR (typically $>100$) template spectra covering a wide range of effective temperatures.  Here, we explore temperature changes for G, K and M main-sequence dwarfs (6000\,K and cooler), but the technique could be expanded to any temperature range and/or surface gravity, provided that one has access to a similar library of template spectra.\\

Measurements of \dtemp for G, K and M main-sequence stars can be performed within LBL by requesting a given combination of temperature gradient and rotational broadening.  \dtemp measurements, along with their associated uncertainties, are computed simultaneously with velocity measurements and returned in single table.

\section{Template construction\label{sect:templates}}
In this demonstration of the differential temperature measurement, we focus on datasets obtained with SPIRou \citep{donati_spirou_2020} and HARPS \citep{pepe_harps:_2002,mayor_setting_2003} instruments to showcase the performances of this activity indicator at both optical and infrared wavelengths. M dwarf templates from SPIRou that were used are observations from the SPIRou legacy survey \citep{fouque_spirou_2023, moutou_characterizing_2023} while K and G stars templates are archival observations obtained in the context of exoplanet transit surveys. For SPIRou, these templates are byproducts of the APERO \citep{cook_apero_2022} data reduction software. For HARPS, we performed a bulk archival query from the ESO archive \citep{delmotte_eso_2006} for all objects within 50\,pc having $>$100 visits. HARPS spectra were telluric-corrected with the correction routine available within the LBL package \citep{artigau_line-by-line_2022}, which nulls the correlation function of telluric absorbers by adjusting the optical depth of telluric absorption from models \citep{bertaux_tapas_2014}. This technique closely matches the mathematical framework by \cite{allart_automatic_2022} with the difference that one adjusts an optical density rather than an effective pressure and optical density.

Line profiles are, overall, affected by 4 main parameters; stellar temperature, metallicity, surface gravity and stellar rotation. As we are interested only in the temperature dependency of our templates, we selected only slow rotators from all HARPS and SPIRou templates. {From the full-width at half-maximum of the cross-correlation function of our templates, we estimate that all objects have a vsini$<3$\,km/s, but note that constraints on the vsini significantly below the combined contribution of the instrument resolution (2.6\,km/s and 4.3\,km/s for HARPS and SPIRou) and intrinsic stellar line widths is challenging}. Selecting objects with a uniform metallicity distribution is more challenging. Metallicity measurements are notoriously challenging for M dwarfs (e.g., \cite{jahandar_comprehensive_2023} and references therein) and all references overlapped with only a handful of targets within our template list. Rather than using a very heterogeneous set of references, some with contradictory values for the same targets, we opted to use the metallicity measurements from the GAIA DR3 catalog (Vizier catalog \textsc{I/355/paramp}; \cite{gaia_collaboration_vizier_2022}) that provides, among other parameters, an estimate of the temperature, surface gravity and metallicity for most of our template stars. Stars for the current analysis were kept if they met the following criteria :

\begin{itemize}
    \item Have parameters defined in the Gaia DR3 catalog
    \item A temperature between 2500 and 6500\,K
    \item A metallicity comprised between $-0.5$ and $+0.5$\,dex
    \item A surface gravity consistent with the main sequence. From the GAIA catalog, $\log\,$g decreases monotonically from 4.9 to 4.2 respectively for T$_{\rm eff}$ of 3000 to 6000. We kept objects within 0.2\,dex of this relation.    
\end{itemize}

For all the analysis below, the temperatures used were drawn from this catalog. While not necessarily the most precise, it has the advantage of being homogeneous in its biases for the entire sample. 

\begin{figure}[!htbp]
    \centering
    \includegraphics[width=\linewidth]{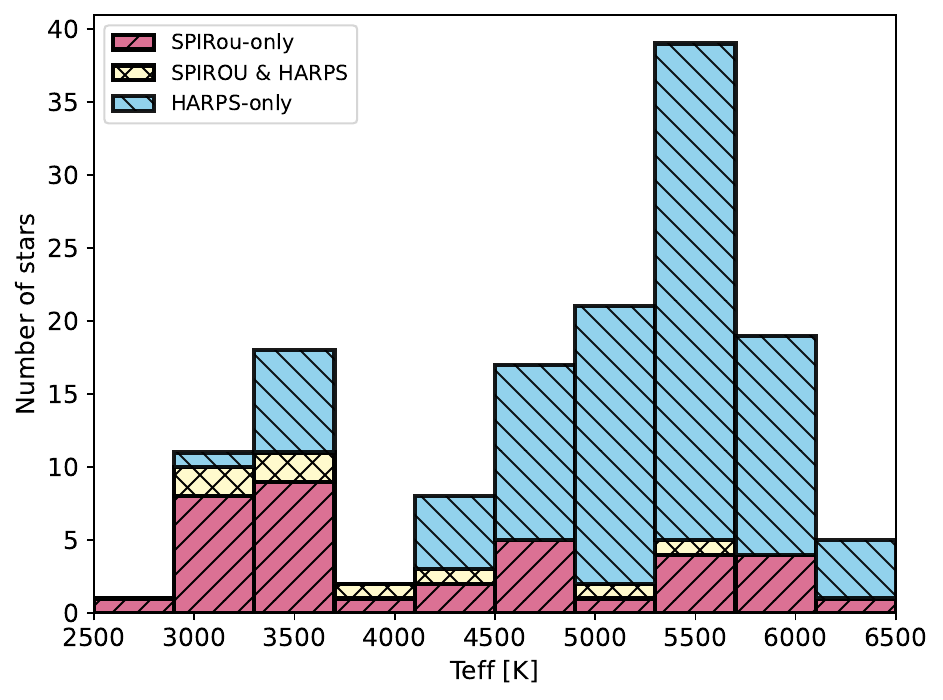}
    \caption{Temperature distribution of the stars used for the SPIRou and HARPS templates. A total of \nstarsspirou\ stars were used for SPIRou, \nstarsharps\ with HARPS, {among which \nstarsoverlap\ are in common.}}
    \label{fig:histogram}
\end{figure}

Table~\ref{tbl:models} compiles all stars used as templates, the instrument(s) concerned, the effective temperature and the systemic velocity used for spectra registering. The temperature distribution for the SPIRou and HARPS templates is shown in Figure~\ref{fig:histogram}. The M dwarf domain (2800--4000\,K) is well covered for SPIRou while mid-K dwarfs to G stars are very well covered for HARPS with a sparser sampling of mid-to-late-M dwarfs ($<$3200\,K). Each template is constructed from the median combination of all observations of each target.

\begin{figure*}[!htbp]
    \centering
    \includegraphics[width=0.495\linewidth]{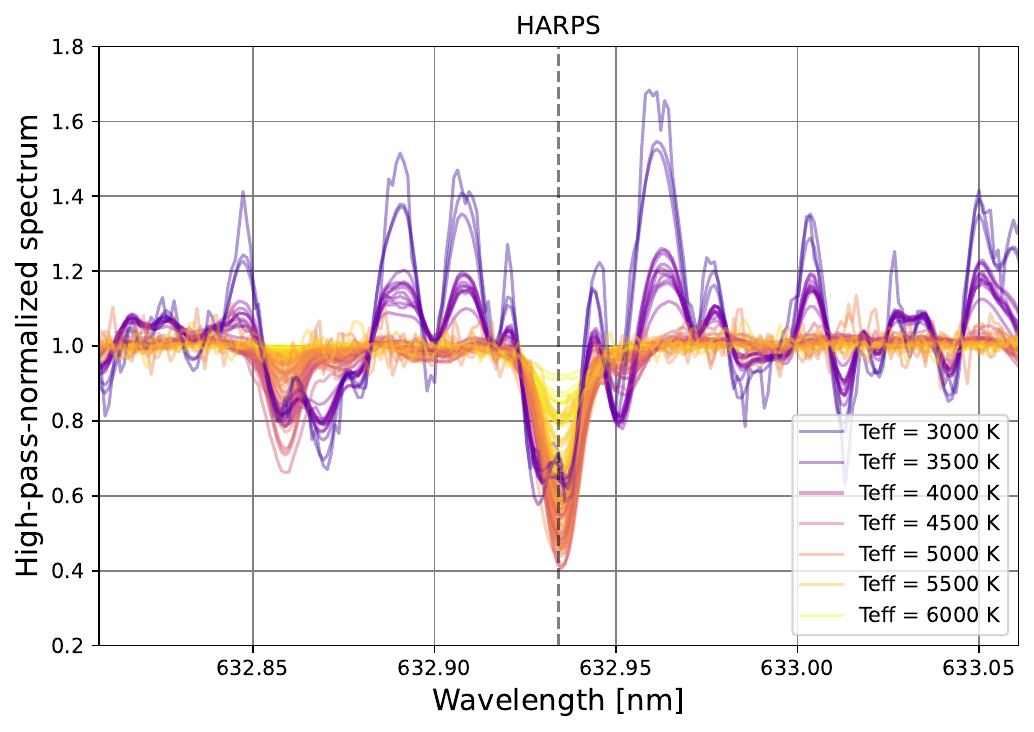}
    \includegraphics[width=0.495\linewidth]{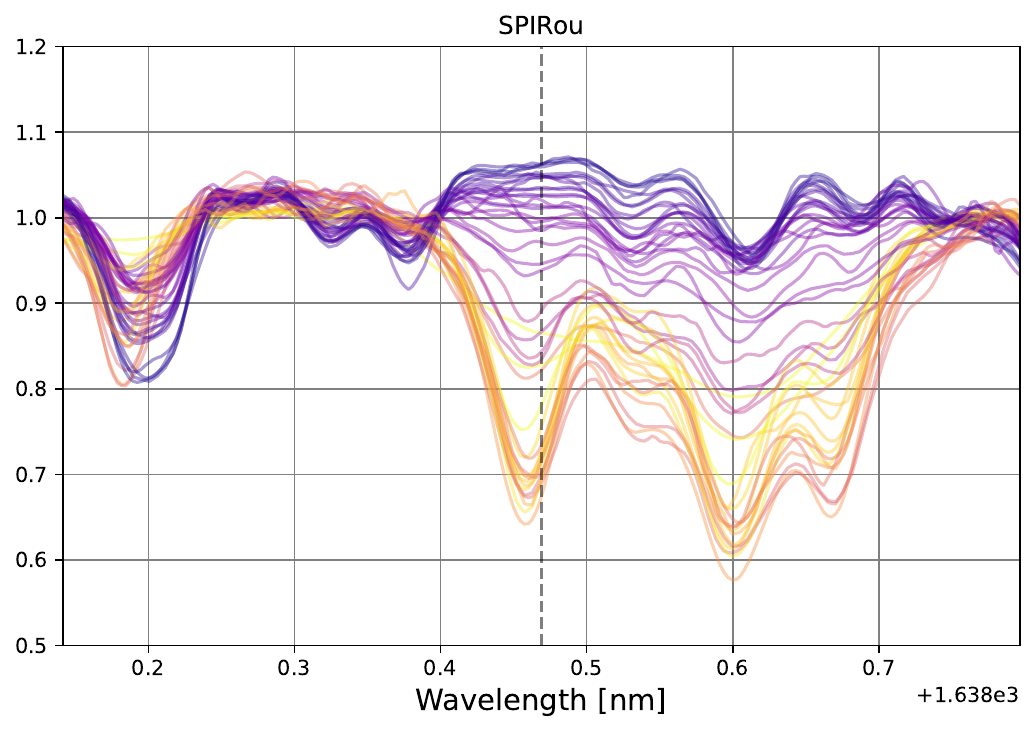}

    \caption{Sample spectroscopic features in the HARPS (left) and SPIRou (right) for sample domains. The temperature dependency of the spectral element highlighted with a dashed line is shown in Figure~\ref{fig:line_spirou_harps_2}. The line centered at 632.935\,nm is a Ni\,I feature while the feature at 1638.50\,nm is tentatively attributed to an Fe\,II line.}
    \label{fig:line_spirou_harps_1}
\end{figure*}

\begin{figure}[!htbp]
    \centering
    \includegraphics[width=0.9\linewidth]{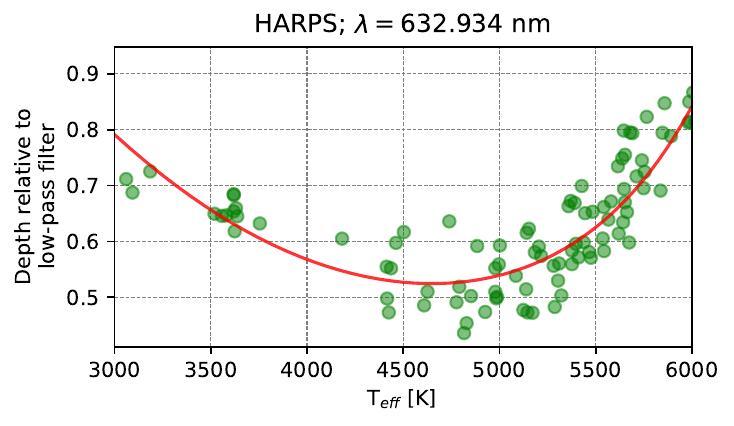}
    \includegraphics[width=0.9\linewidth]{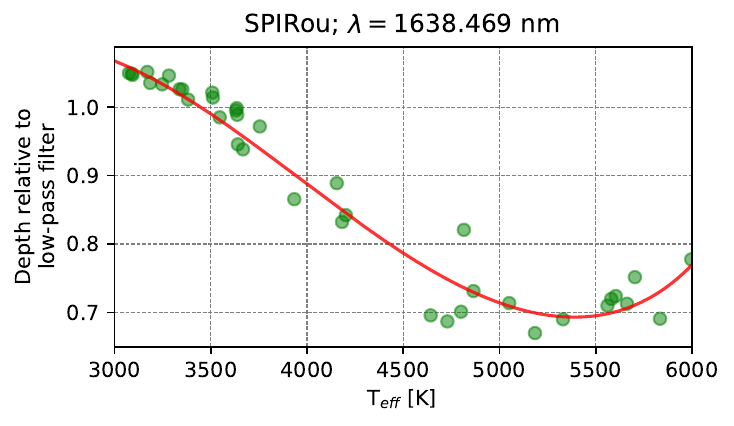}
    \caption{ 
Temperature dependency of the spectral element shown in Figure~\ref{fig:line_spirou_harps_1} with the 4$^{\rm th}$-order polynomial fit for that pixel. Residuals to the fit have an RMS of \RMSresidualharps\ and \RMSresidualspirou\ for SPIRou and HARPS. 
}
    \label{fig:line_spirou_harps_2}
\end{figure}

All spectra were registered to a common systemic velocity and high-passed with a median filter on a scale of 100\,km/s. As illustrated in Figure~\ref{fig:line_spirou_harps_1} for sample domains, one sees a coherent evolution of the depth of spectral features with the temperature of the templates. For a given spectral pixel (see Figure~\ref{fig:line_spirou_harps_2}), the relative flux depth can be fitted with a low-order polynomial as a function of temperature.

For the 2500-6000\,K domain, we empirically found that a 4$^{\rm th}$-order polynomial with temperature is sufficient to fit our data properly. The analytical derivative of this polynomial gives its derivative at any given temperature. This provides the $\partial A_0 / \partial T$ needed in equations~\ref{eq2} through \ref{sum_rv}.

\begin{figure}[!htbp]
    \centering
    \includegraphics[width=\linewidth]{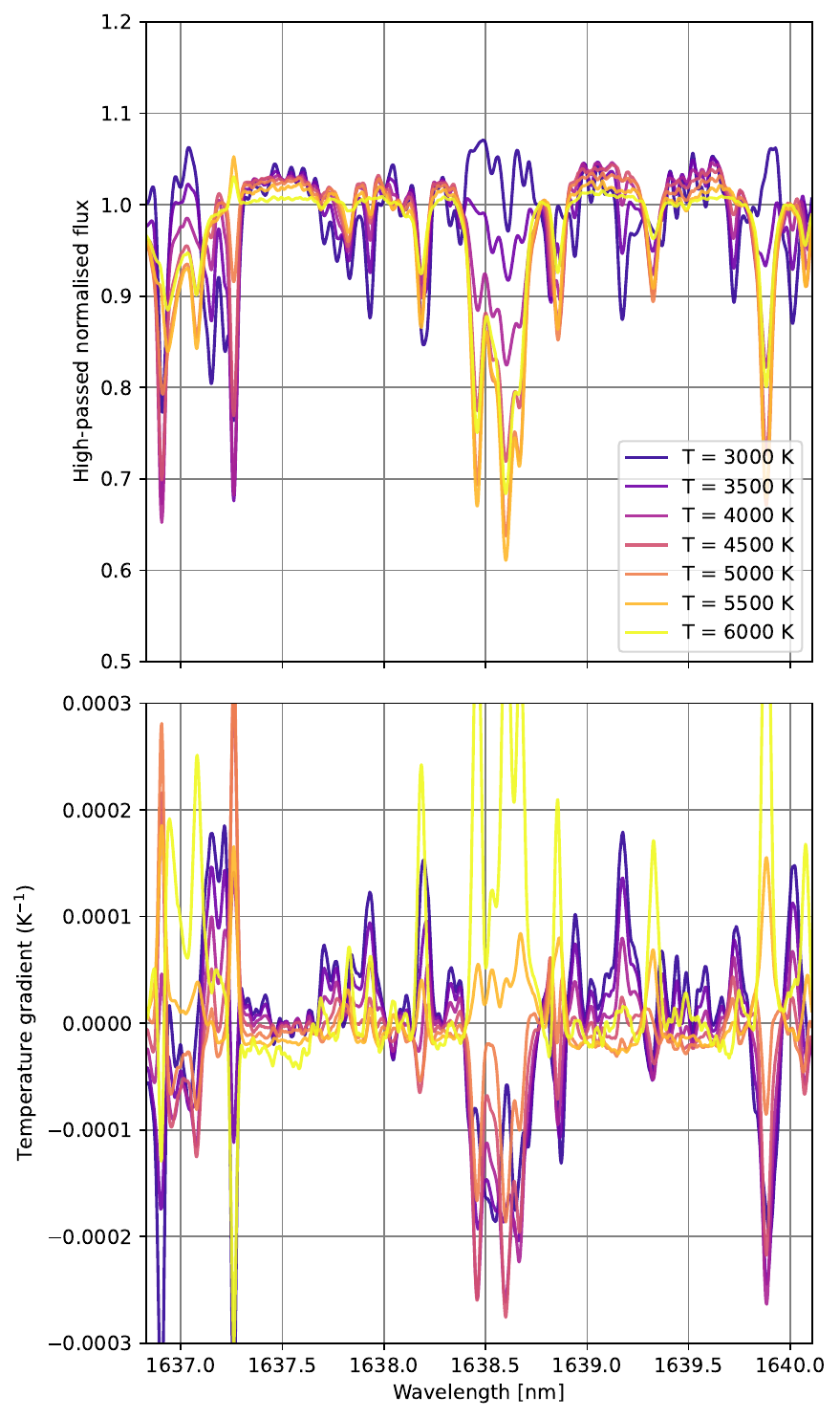}
    \caption{[Top] Sample average spectra at 500\,K steps for SPIRou. The corresponding gradient spectra are shown on the bottom row. One can see that lines that change in depth in the top plot correspond to strong temperature gradient features in the bottom plot. }
    \label{fig:gradient_spirou}
\end{figure}

Figure~\ref{fig:line_spirou_harps_2} 
illustrates the $\partial A_0 /\partial T$ spectra over the 3000 to 6000\,K domain for temperature steps of 500\,K for the sample domains shown in Figures~\ref{fig:line_spirou_harps_1} and \ref{fig:line_spirou_harps_2} for the SPIRou spectra. The spectrum is expressed in K$^{-1}$, consisting of a fractional variation in the depth of a feature expressed as a fraction of the local low-pass filtered pseudo-continuum. The amplitude of the temperature gradient spectra is generally very low (typically less than a few $10^{-4}$\,K$^{-1}$ (i.e., at most a few tens of percent in fractional line depth change for a 1000\,K change in temperature). {Figure~\ref{fig:gradient_spirou} illustrates the mean templates and the corresponding temperature gradients for a representative domain within the SPIRou domain.}

The scatter to the fit shown in Figure~\ref{fig:line_spirou_harps_2} informs on the precision one needs on the stellar temperatures. Given the RMS to the polynomial fit with temperature (See Figure~\ref{fig:line_spirou_harps_2}), considering that typical spectral gradients are on the order of $10^{-4}$\,K$^{-1}$, the required precision for the stellar temperatures such that is a minor contributor to the dispersion is $100-300$\,K (i.e., with the gradients considered, a 100\,K error contributes, in quadrature, 0.01 to the scatter). As the typical error in temperature for the GAIA DR3 values, compared to detailed spectroscopic analysis, is 110\,K \citep{andrae_gaia_2023}, we conclude that this is a relatively minor contributor to the overall scatter in the relations shown earlier. Going into a detailed, target-per-target $T_{\rm eff}$ determination through spectroscopic analysis would not, in the present context, change our results.

We note that the concept of measuring precise differential temperature using per-line depth measurements is not new. {\cite{gray_precise_1991} and \cite{gray_spectral_1994}} used line-depth ratios to measure temperatures of late-F to early-K stars down to differential precision of $\sim$10\,K by selecting iron and vanadium that are highly sensitive to temperature changes within the $r$ photometric bandpass. They used an empirical line-depth ratio relation with color, as a proxy of temperature, to derive a polynomial relation that is similar conceptually to the generalized per-spectral-pixel polynomial relations presented here (See in particular Figures~1 and 2 in \cite{gray_precise_1991}). One key difference with this early work is that we exploit the full information content of the spectra and use a purely empirical determination of the temperature changes in spectra without prior knowledge of the dominant absorber at each wavelength. This is particularly important for M dwarf spectra that are dominated by molecular bands that are notoriously challenging to reproduce, at the spectral resolution considered here, with stellar models {\citep[e.g.,][]{artigau_optical_2018, jahandar_comprehensive_2023}.
}

\section{Application to pRV datasets\label{sect:prv}}

We used the framework described above and applied it to three representative stars with the goal of constraining their rotation period through the measured temperature variations induced by stellar heterogeneities:
\begin{itemize}
\item AU Mic\footnote{CFHT run IDs : \RUNIDAUMIC.}: a very active M dwarf at infrared wavelengths.
    \item Barnard's star\footnote{CFHT run IDs: \RUNIDBARNARD.}: an RV-quiet M dwarf at infrared wavelengths.
    \item \epse\footnote{ESO program ID: \RUNIDepseri.}: an active K dwarf at optical wavelengths.
\end{itemize}

\subsection{Active M dwarf, near-infrared: AU Mic \label{sect:aumic}}
\begin{figure*}[!tbp]
    \centering
    \includegraphics[width=0.99\linewidth]{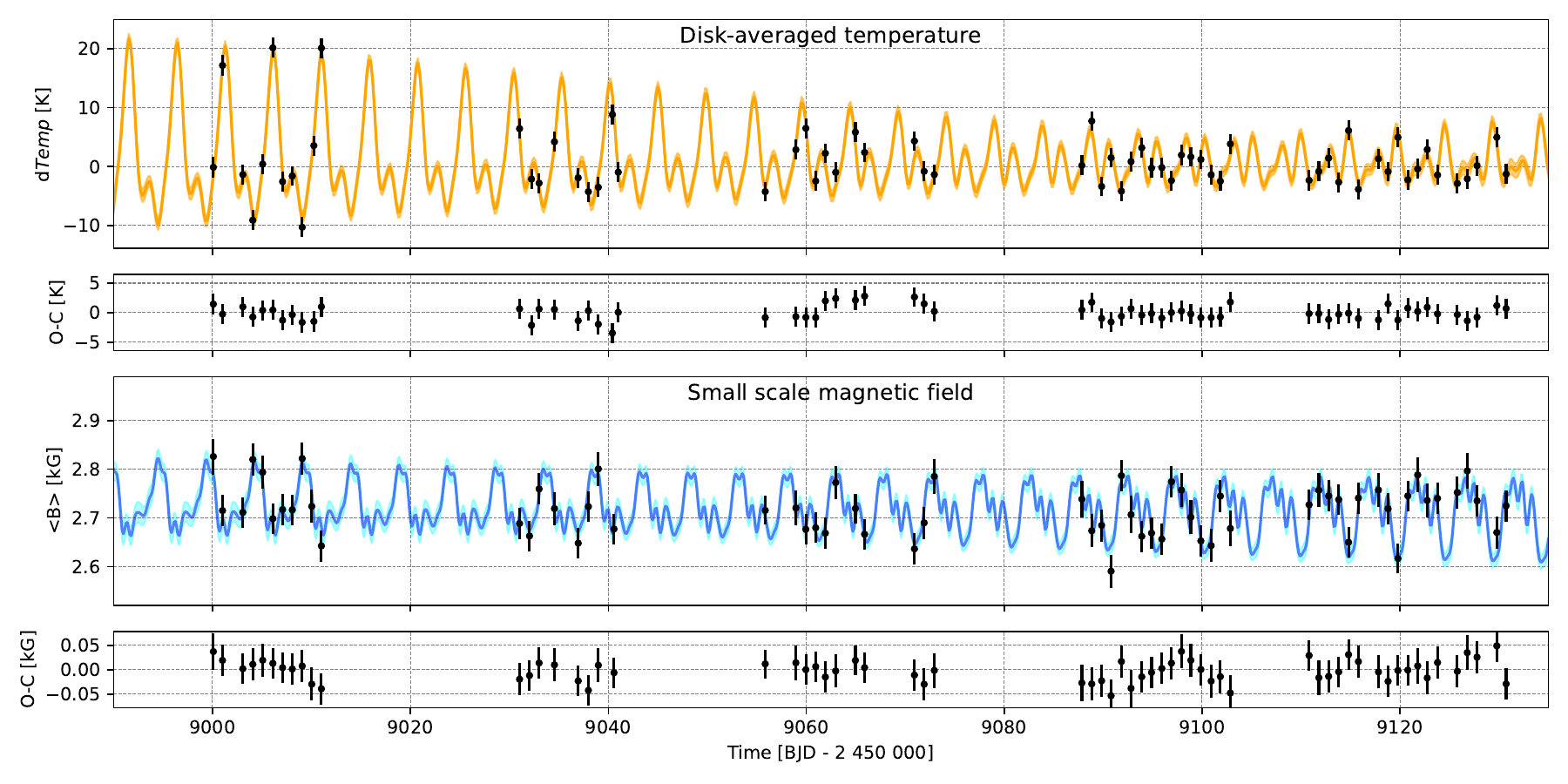}
    \caption{Sample sequence within the AU Mic \dtemp time series. A Gaussian-process fit to the data is shown as an overplot (top panel), with the residual to that fit in the bottom panel. One sees the envelope of the signal decaying in strength through the first half of the sequence as well as the notable presence of a strong second harmonic signal (secondary bump appearing around JD=2\,459\,030). The small-scale magnetic field measurements form \cite{donati_magnetic_2023} are shown for comparison. \dtemp and \magb show a remarkable agreement in their overall behavior, with a strong anticorrelation.}
    \label{fig:gp_aumic}
\end{figure*}

\begin{figure*}[!tbp]
    \centering
    \includegraphics[width=0.99\linewidth]{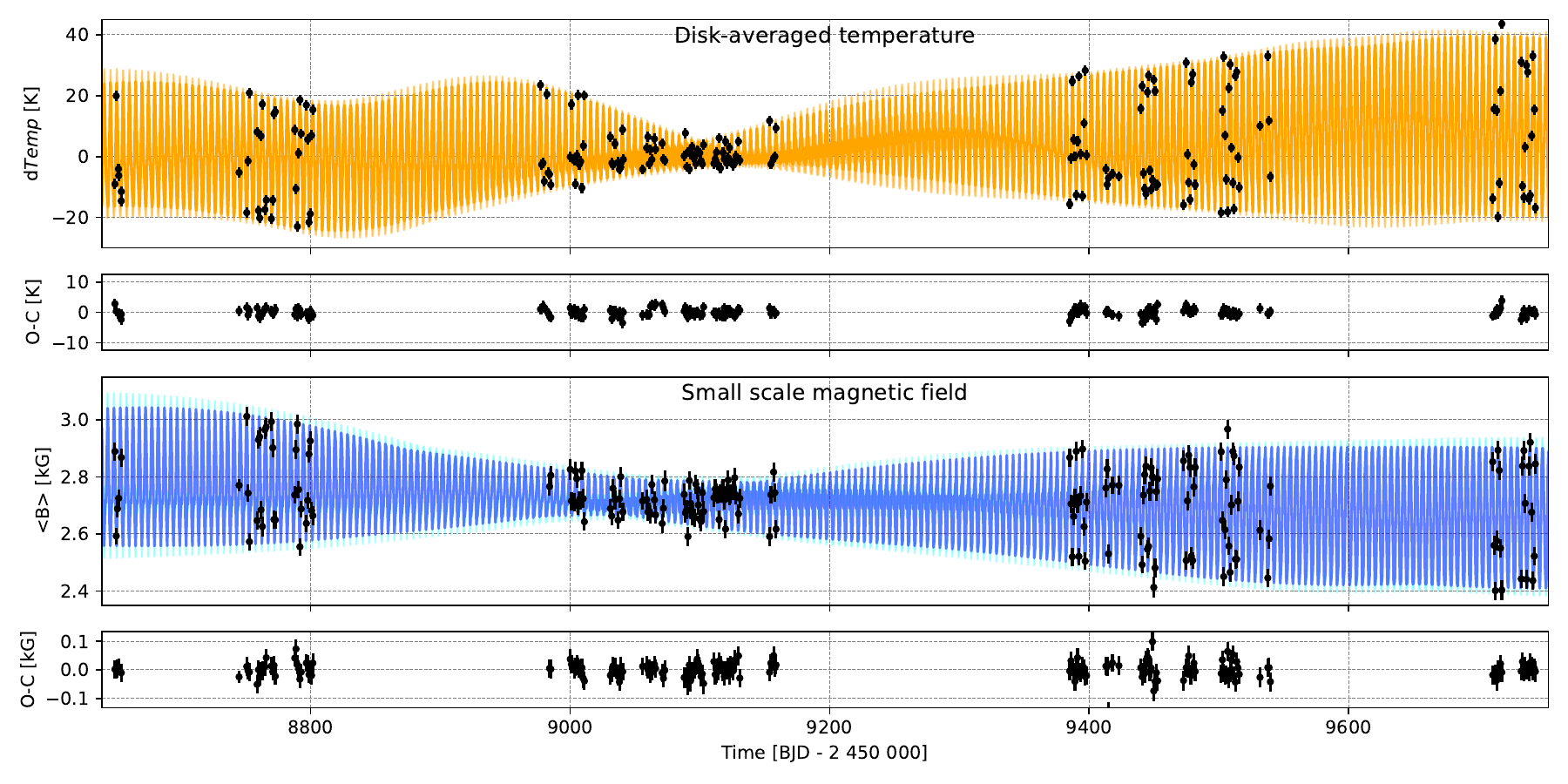}
    \caption{Same as Figure~\ref{fig:gp_aumic} over the entire time series.}
    \label{fig:gp_aumic2}
\end{figure*}

AU\,Mic is among the best examples of active M dwarfs in the solar vicinity, with a dust disk \citep{kalas_discovery_2004}. The star has two known transiting planets \citep{plavchan_planet_2020, martioli_new_2021, szabo_changing_2021, szabo_transit_2022}, with the detection of a third planet through TTV \citep{wittrock_validating_2023} and a tentative detection of a fourth one from radial velocity monitoring \citep{donati_magnetic_2023, wittrock_validating_2023}. The dataset analyzed here is the same as presented in \cite{donati_magnetic_2023}, and has been reduced with \textsc{APERO} version~{0.7.284}. In our analysis we used the temperature gradient for 3500\,K convolved with a rotational kernel of 9\,km/s. The resulting \dtemp time series is shown in Figures~\ref{fig:gp_aumic}  and \ref{fig:gp_aumic2}. The \dtemp time series shows a strong peak-to-peak variation at the level of $\sim$40\,K.

The time series was fitted with a Gaussian process (GP) using a quasi-periodic covariance function (\citealt{haywood_planets_2014, rajpaul_gaussian_2015, stock_gaussian_2023}):
\begin{equation}
k(\tau) = \sigma^2_{\rm GP} \exp  \left(-\frac{\tau^2}{2l^2} - \Gamma \sin^2 \left( \frac{\pi\tau}{P_{\rm rot}} \right)  \right) + \sigma_{\rm jit}^2
\label{eq6}
\end{equation}

As defined in \cite{stock_gaussian_2023}, the parameters are described here accordingly:

\begin{itemize}
    \item $\sigma_{\rm GP}$ is the amplitude of the GP component given in Kelvin.
    \item $l$ is the correlation length scale, in units of days, of the GP squared-exponential component, it represents the coherence timescale of repeating structures in the periodic signal.
    \item $\Gamma$ defines the dimensionless relative weight between the GP sine-squared component and squared-exponential component.
    \item $P_{\rm rot}$ is the period of the GP sine-squared component in days.
    \item $\tau$ is the time lag, in days, between observations for which we are determining the covariance 
    \item $\sigma_{\rm jit}$ is the excess noise (jitter term; in Kelvin) to fully account for the variance in our data. A value of $\sigma_{\rm jit}$ converging to zero implies that the variance in the data is fully described by the GP model and the propagated uncertainties in the data.
\end{itemize}

The recovered GP model parameters and the corresponding constraints on the physical parameters of the evolution of the \dtemp time series of AU Mic are given in Table~\ref{tbl:aumic}. The time series of \dtemp and the small-scale magnetic field ($<$$B$$>$) of AU Mic taken from \cite{donati_magnetic_2023} are shown in Figure~\ref{fig:gp_aumic} (subset of 130\,days for clarity), with the residual to the corresponding GP fit. The full time series is shown in Figure~\ref{fig:gp_aumic2}. The derived rotation period of \postLINAUMiclnP\,days is in agreement with the \postLINAUMicBmaglnP\,days for the posterior value obtained with a GP fit on \magb in \cite{donati_magnetic_2023} (Table~2 therein). 

\begin{figure}[!htbp]
    \centering
z    \includegraphics[width=0.99\linewidth]{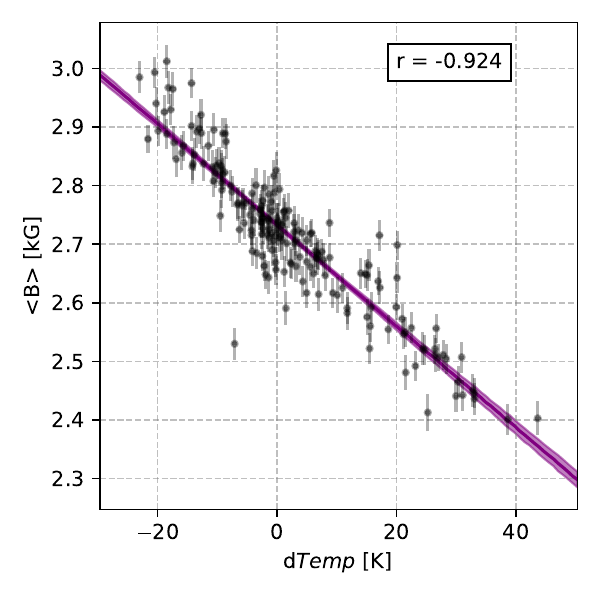}
    \caption{\magb as a function of \dtemp for the time series shown in Figure~\ref{fig:gp_aumic}. An increase in temperature is associated with a decrease in the small-scale magnetic field. This is qualitatively understood as a decrease in the number of active regions on the observed hemisphere, leading to a smaller level of activity, fewer observed stellar spots and, hence, a hotter average temperature. The correlation between the quantities (\pearsondtempbmag) is excellent, suggesting that they trace the same underlying physical processes. The 1-$\sigma$ envelope of Equation~\ref{eq:bmag} is highlighted in purple.}
    \label{fig:gp_aumic_correlation}
\end{figure}

As shown in Figure~\ref{fig:gp_aumic_correlation}, the \dtemp time series is remarkably anti-correlated with the small-scale magnetic field, with a Pearson-r correlation between \dtemp and \magb of \pearsondtempbmag, indicating that the two draw their origin in the same underlying physical processes. Stronger magnetic fields are linked to an apparent decrease in the bulk temperature of AU\,Mic.

This anti-correlation provides an effective means of determining the small-scale magnetic field \magb of any M dwarfs through a \dtemp measurement derived from a high-resolution spectrograph. From  Figure~\ref{fig:gp_aumic_correlation}, we derive the following empirical calibration for AU Mic between \magb and \dtemp (respectively in kGauss and Kelvin), accounting for the uncertainties in both \magb and \dtemp (i.e., using an orthogonal distance regression; \cite{brown_statistical_1990}):

\begin{equation}
\label{eq:bmag}
\slopedtempbmagfit
\end{equation}

The above relation has a typical dispersion of 0.047\,kG compared to a median error on \magb of 0.027\,kG. While the anti-correlation is remarkable in the case of AU Mic, it remains to be seen if it is also present for stars that are much less active and at different temperatures; we defer this question for future work. Considering that \magb is an excellent proxy for RV jitter \citep{haywood_unsigned_2022} and that this quantity displays, at the very least in the case of AU Mic, a strong anti-correlation of \dtemp, we expect that \dtemp will be similarly good for RV activity filtering.

\begin{table}[!htbp]
\setlength{\tabcolsep}{2pt} 
\caption{Priors and posteriors on the GP model of AU\,Mic. $\mathcal{LU}\left(a,b \right)$ refers to the the log-uniform (``Jeffreys'') distribution.}\label{tbl:aumic}
\begin{tabular}{lcr}
\hline
 Parameter & Prior &  Posterior\\\hline
$\sigma_{\rm GP}$ (K) & \priorAUMiclnA & \postLINAUMiclnA \\
$l$ (days) & \priorAUMiclnl & \postLINAUMiclnl \\
$\Gamma$ & \priorAUMiclngamma & \postLINAUMiclngamma \\
$P_{\rm rot}$ (days) & \priorAUMiclnP & \postLINAUMiclnP \\
$\sigma_{\rm jit}$ (K) & \priorAUMiclns & \postLINAUMiclns \\
\hline
\end{tabular}
\end{table}

\subsection{Inactive M dwarf, near-infrared: Barnard's star}
Barnard's star {\citep{barnard_small_1918}} is the second nearest stellar system to the Sun at a distance of  $1.8282\pm0.00013$\,pc \citep{gaia_collaboration_gaia_2021}, and has been the topic of much interest in exoplanet searches. Despite numerous attempts at finding planets, all discovery claims to date have been challenged, from early astrometric claims \citep{van_de_kamp_astrometric_1963} to the more recent claim by \cite{ribas_candidate_2018} of a planet with a 233-day orbital period. 

\begin{figure*}[!htbp]
    \centering
    \includegraphics[width=0.99\linewidth]{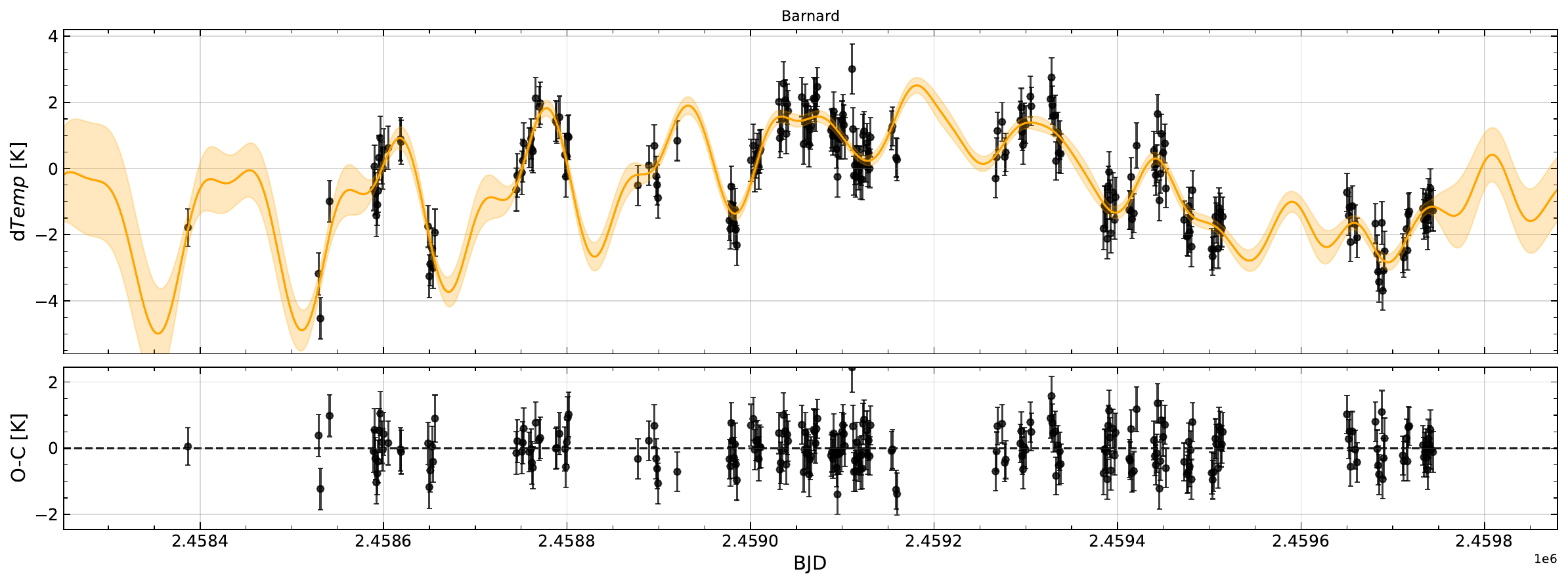}
    \caption{\dtemp time series for Barnard's star [top]. As expected from the modest level of activity of this thick disk star, a much milder variability is observed than for AU\,Mic, but nonetheless one notes the rotation-induced variability. The residual to the GP fit [bottom] shows sub-K RMS, consistent with the $\sigma_{\rm jit} =$\postLINBarnardlns\,K reported in Table~\ref{tbl:gl699}.}
    \label{fig:gp_gl699}
\end{figure*}

Using the dataset presented in \cite{artigau_line-by-line_2022}, we compute the temperature variations of Barnard's star, which is notable for its low level of activity and, correspondingly, low level of RV jitter. We did not convolved the derivative spectrum as Barnard's star is a very slow rotator with an equatorial velocity of $<$100\,m/s. Contrary to AU\,Mic, one expects a much more modest activity signal. Using the same GP kernel as in Section~\ref{sect:aumic} we constrain the time covariance of the time series. Figure~\ref{fig:gp_gl699} shows the evolution of Barnard's star temperature over nearly 4 years. Its temperature fluctuates by $\pm2$\,K peak-to-peak with a semi-periodic behavior. The GP fit parameters are provided in Table~\ref{tbl:gl699}.

The rotation period recovered here (\postLINBarnardlnP\,days) is close to the published value inferred through activity-rotation relations {\citep[148.6\,days;][]{suarez_mascareno_rotation_2015}} and spectropolarimetry \citep[$137\pm5$\,days;][]{fouque_spirou_2023}). The value is mildly discrepant with the photometric measurements of 130.4\,days obtained by \cite{benedict_photometry_1998} with HST fine-guider photometry; however, we note that this reference does not report uncertainties. Furthermore, M dwarfs may very well experience differential rotation, therefore, features inducing photometric modulations in \cite{benedict_photometry_1998} may not occur at the same latitude as structures that led to other rotation measurements. The uncertainties on GP parameters from \dtemp are comparable to those derived in \cite{fouque_spirou_2023} from longitudinal magnetic field measurements.

\begin{table}[!htbp]
\setlength{\tabcolsep}{2pt} 
\caption{Barnard's star priors, posteriors, and constraints on the GP model.}
\label{tbl:gl699}
\begin{tabular}{lcr}
\hline
 Parameter & Prior &  Posterior\\\hline
$\sigma_{\rm GP}$ (K) & \priorBarnardlnA & \postLINBarnardlnA \\
$l$ (days) & \priorBarnardlnl & \postLINBarnardlnl \\
$\Gamma$ & \priorBarnardlngamma & \postLINBarnardlngamma \\
$P_{\rm rot}$ (days) & \priorBarnardlnP & \postLINBarnardlnP \\
$\sigma_{\rm jit}$ (K) & \priorBarnardlns & \postLINBarnardlns \\
\hline
\end{tabular}
\end{table}

\subsection{\epse}
\epse is the second closest K dwarf to the Sun after $\alpha$\,Centauri\,B and the closest isolated one. It is located $3.220\pm0.004$\,pc \citep{gaia_collaboration_gaia_2021} from the Sun and has had its radius and temperature, $0.74\pm0.01$\,R$_\odot$ and $5039\pm126$\,K, determined through interferometric measurements \citep{baines_confirming_2012}. RV measurements point to a planetary-mass companion of a $\sim$7-year orbit  but pRV detection of planets around \epse is challenging due to the relatively high level of activity of this 0.4--0.8\,Gyr-old star \citep{mamajek_improved_2008} and orbital parameters (planet mass and period) are inconsistent between various recent studies of the system \citep{hatzes_evidence_2000, feng_revised_2023, llop-sayson_constraining_2021}. This star shows activity-induced jitter at the $\sim$10\,m/s level (\citealt{giguere_combined_2016, hempelmann_measuring_2016, lanza_measuring_2014}) which is comparable to the RV semi-amplitude of \epse b {\citep{feng_revised_2023}}. While activity can be filtered to  2-3\,m/s \citep{giguere_combined_2016, roettenbacher_expres_2022}, this has only been demonstrated on timescales commensurate with the rotation period and not on a timescale similar to the activity cycle of K dwarfs.

\begin{figure*}[!htbp]
    \centering
    \includegraphics[width=0.99\linewidth]{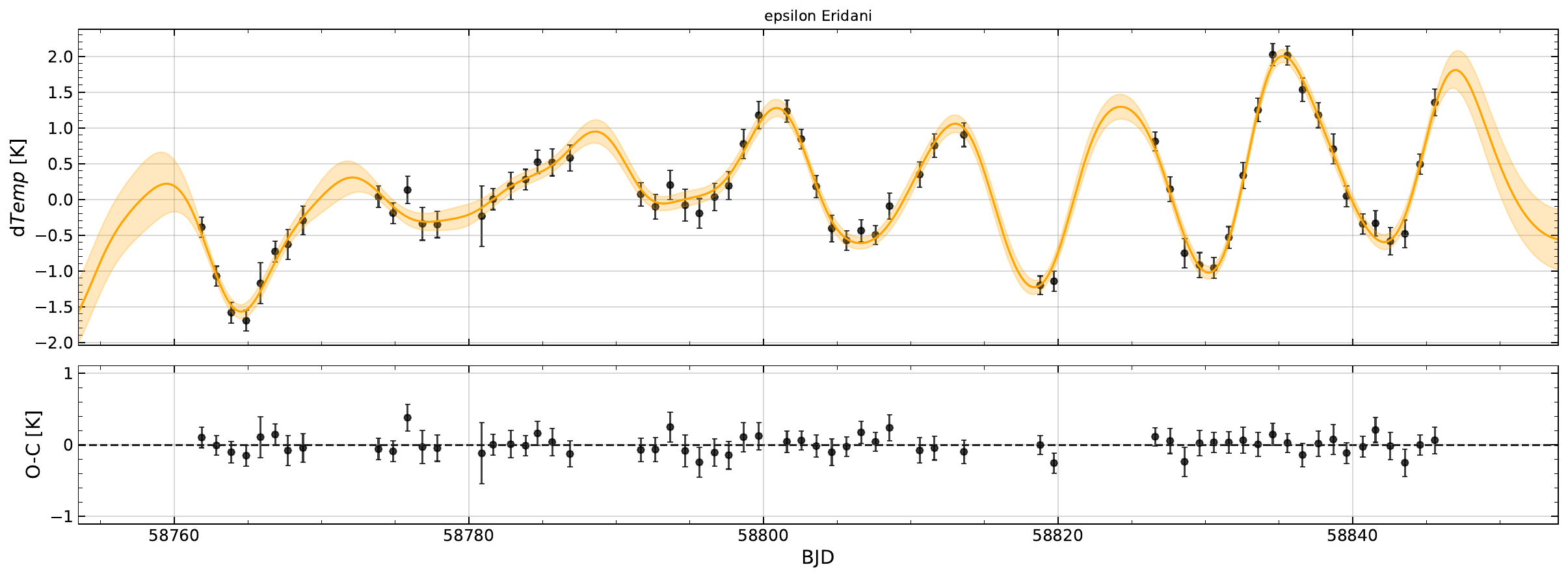}
    \caption{\dtemp time series for \epse at optical wavelength with HARPS. Rotation-induced variability is detected in the second half of the dataset and is readily described by a semi-periodic GP fit [top]. The residuals to the GP fit are at the 0.1\,K level [bottom]. In both time series, individual points have been binned on a nightly basis.
    \label{fig:epseri}}
\end{figure*}

\epse has been observed with HARPS 741 times over the 2003-2019 period, but we focus our analysis on the observations taken between 4 October 2019 and 27 December 2019. Over this period of 83 days, it was observed on 62 nights with, on most occasions, 3 spectra per night (190 spectra total). This is the densest monitoring of this star and it is the best dataset to constrain the evolution of its \dtemp time series. The template used in the \dtemp computation was constructed from all observations but only the 2019 sequence is sufficiently dense for proper retrieval of the rotation period. For our analysis we used a temperature gradient spectrum corresponding to 5000\,K convolved with a 3\,km/s rotation kernel \citep{giguere_combined_2016}. The \dtemp time series of \epse\, shown in Figure~\ref{fig:epseri}, allow for an accurate GP retrieval of the rotation period; the properties of the fit are reported in Table~\ref{tbl:epseri}.

Notable among main sequences stars in the solar neighborhood, \epse is sufficiently close to directly resolve its active regions through interferometry, and the correspondence between spatially resolved star spot and RV jitter measured with an optical pRV spectrograph (EXPRES) is discussed in \citet{roettenbacher_expres_2022}. Performing a joint analysis of the \dtemp with interferometric data would be even more compelling as it would lift degeneracies between filling factors and starspot temperature contrast.

The GP fit to the \epse\ time series differs from those derived through optical \& near-infrared high-resolution spectropolarimetry and TESS photometry \citep{petit_multi-instrumental_2021}. The rotation period derived is \hbox{$P_{\rm eq} = 10.77 \pm 0.06$\,days} at the equator (see Figure~9 therein) with a differential rotation \hbox{$d/\Omega = 0.11 \pm 0.01$\,rad/d}. The equatorial velocity is discrepant with our rotation period \postLINepsilonEridanilnP\,days, but this discrepancy can be readily explained if the dominant features creating the modulation shown in Figure~\ref{fig:epseri} are not equatorial.  {\cite{petit_multi-instrumental_2021}} derives a polar rotation of 13.3\,days, and using Equation~6 therein, we determine that the temperature differences we have measured occur at a latitude of \omegaepseri.  This is at higher latitudes than for the Sun (typically within latitudes of $\pm30^\circ$) but reported for some Kepler quarters for the Sun-like star HD~173701 through asteroseismology \citep{bazot_butterfly_2018} and similar to values derived for the K1V star HD 106225 \citep{hatzes_spot_1998}.

\begin{table}[!htbp]
\setlength{\tabcolsep}{2pt} 
\caption{\epse priors, posteriors and constraints on the GP model.}
\label{tbl:epseri}
\begin{tabular}{lcr}
\hline
 Parameter & Prior &  Posterior\\\hline
$\sigma_{\rm GP}$ (K) & \priorepsilonEridanilnA & \postLINepsilonEridanilnA \\
$l$ (days) & \priorepsilonEridanilnl & \postLINepsilonEridanilnl \\
$\Gamma$ & \priorepsilonEridanilngamma & \postLINepsilonEridanilngamma \\
$P_{\rm rot}$ (days) & \priorepsilonEridanilnP & \postLINepsilonEridanilnP \\
$\sigma_{\rm jit}$ (K) & \priorepsilonEridanilns & \postLINepsilonEridanilns \\
\hline
\end{tabular}
\end{table}

\section{Transit-induced temperature changes\label{sec:titc}}
While variations of the temperature over a full rotation were the focus of the above demonstration datasets, an interesting case of short-term apparent{ disk-averaged} temperature change appears during a transit. As the stellar disk is occulted by a companion (an exoplanet in the case presented here), the averaged temperature will change slightly. Masking the area close to the limb will remove cooler-appearing regions (from the line-of-sight geometry at that position on the stellar surface) and the disk-averaged temperature will therefore increase. Conversely, masking the center of the star, the hottest-appearing area on the star (in the simplistic case without active regions), one will see an apparent cooling of the star. As the transiting planet HD\,189733\,b has been observed several times with SPIRou\footnote{CFHT run IDs : \RUNIDHDxxxxx.}, we predicted the amplitude of this unreported transit effect in our data. The \dtemp was measured with a gradient spectrum at 5000\,K and assuming a 3\,km/s $v \sin i$ 
\citep{cristo_espresso_2024}. The analysis accounts for the Rossiter-McLaughlin (RM) effect \citep{rossiter_detection_1924, mclaughlin_results_1924}; as mentioned in Section~\ref{sect:method}, the \dtemp is computed after determining the average velocity.

\subsection{Constructing a toy model of \dtemp \label{sec:model_dtemp}}
While the limb of a star has a temperature that is cooler than the disk-averaged stellar SED, its exact spectrum is not identical to that of a cooler star. To construct a model prediction of the \dtemp equivalent of the RM effect, we retrieved the PHOENIX Goettigen atmosphere models \citep{husser_new_2013} with their full radiation field. We constructed a 3-D cube with wavelength and spatial position on the stellar disk and constructed a time series of the total spectrum through transit. We assumed no stellar rotation (i.e., stellar spectra are not RV-shifted at the corresponding limb), which is a valid assumption for a slowly rotating star. This leads to a grey lightcurve as the sum of the flux over the entire disk. It represents the classical transit light curve, characterized by the standard central dip centred at mid-transit due to limb darkening of the star (see panel [1] of Figure~\ref{fig:toymodel}). Furthermore, we allow for unocculted regions on the stellar disk, where a given fraction of the star, outside of the transit chord, is covered with regions at different temperatures from the rest of the photosphere. These unocculted regions could be either cooler (spots) or hotter (plages) than the bulk of the photosphere. For unocculted regions, we assumed the same limb-darkening as for the bulk of the photosphere. Figure~\ref{fig:toymodel} illustrates the geometry of the toy-model used here.

To test the accuracy of our methodology, we produced a model lightcurve within the TESS bandpasses and compared it with the TESS phase-folded lightcurve for HD\,189733\,b integrated over the 3D cube, the agreement is excellent, with a $<$0.5\,mmag accuracy on the lightcurve for a $\sim$2.8\% transit depth.

In the same framework, we derived a residual transit spectrum by taking the ratio of the out-of-transit spectrum to that at any transit position. This residual was projected onto the derivative of the spectrum with temperature, determined within the same framework as described in Section~\ref{sect:templates}, but constructed with models at temperatures immediately above and below that of the temperature of interest (in 100\,K steps here). The predicted \dtemp measurement through transit is shown in Figure~\ref{fig:toymodel_predictions}, panel [1], with impact parameters of 0.0, 0.5, 0.69 (nominal value) to 0.9. The impact of the lightsource effect is highlighted in Figure~\ref{fig:toymodel_predictions}, panel [3]. Unocculted regions occupying $2.8$\%  \citep{sing_hubble_2011} of the photosphere and a temperature difference ranging from 0 to 2000\,K. The impact on the \dtemp signal is significant, with an evolution of the morphology of the curve; the central dip strengthens with cooler unocculted regions. The impact on the transit depth, on the contrary, is minimal and would be identical to a change in planet-to-star radius ratio. 

As detailed in Section~\ref{sect:models_vs_templates}, the gradient spectrum derived from models only partially matches the empirically derived one. The toy model constructed here assumes that the atmosphere models are correct in a differential way. Even if all line positions are not accurate, changing the filling factor during transit (e.g.,  lightsource effects in Figure~\ref{fig:toymodel_predictions}) should be correctly predicted, to a first order, provided that gradient spectra used to compute the \dtemp values are derived consistently with the spectra used to construct these effects.

{ \begin{figure}[!htbp]
    \includegraphics[width=\linewidth]{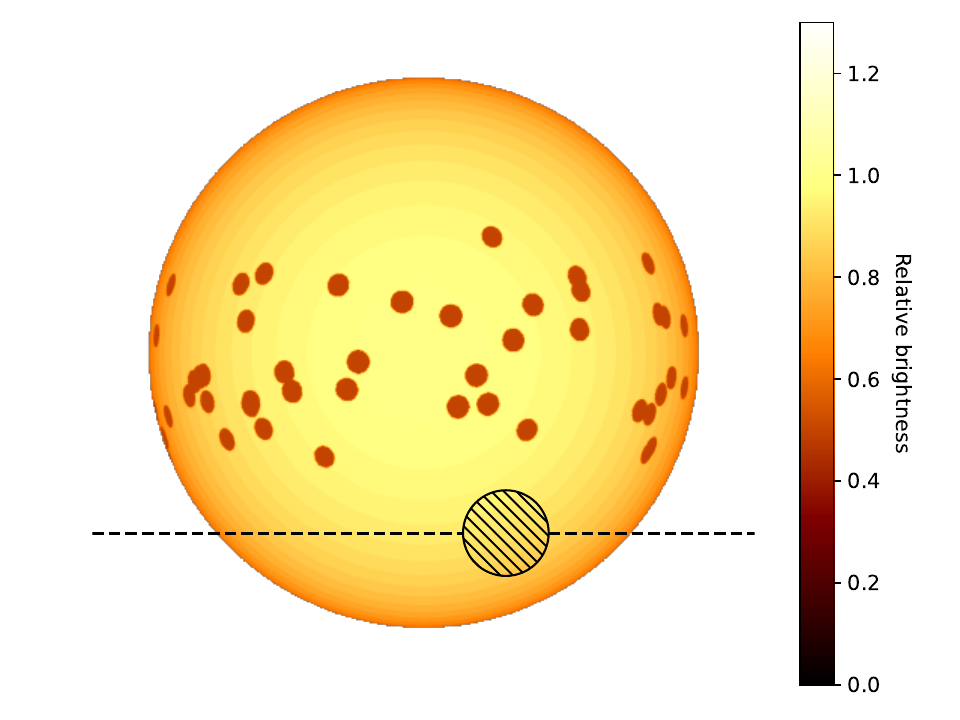}
    \caption{Cartoon representation of the toy model used to predict the amplitude of the \dtemp effect in the HD\,189733\,b transit data. The impact parameter of 0.69 corresponds to that of the system; the relative radii of the host star and planet correspond to literature values. Unocculted spots covering $2.8$\% of the disk are shown (filling factor from \cite{sing_hubble_2011}).}
    \label{fig:toymodel}
\end{figure}}

 \begin{figure*}[!htbp]
    \includegraphics[width=\linewidth]{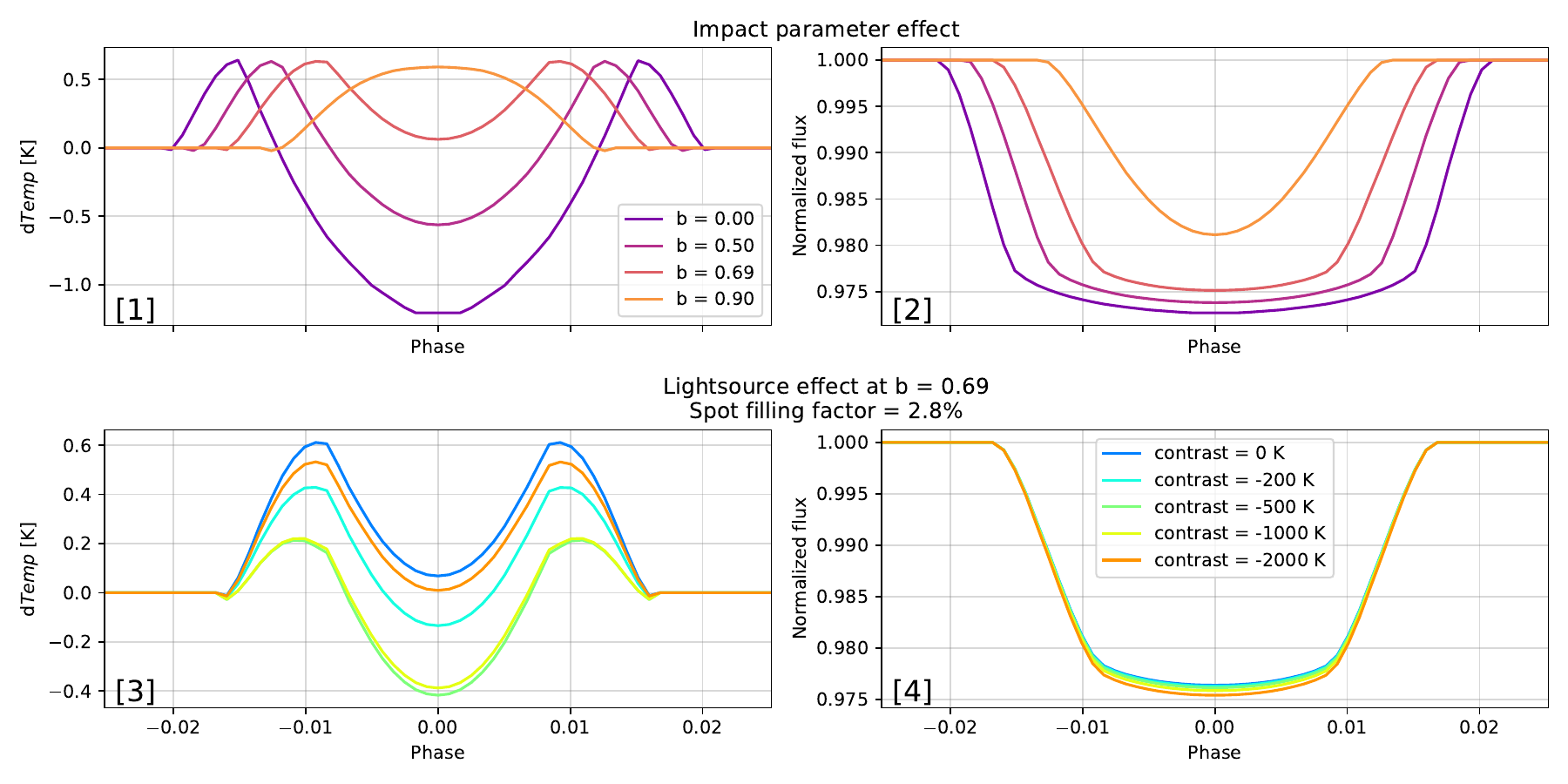}
    \caption{[1] Model \dtemp for different impact parameters for the HD\,189733 system and the corresponding broadband lightcurves [2] integrated over the SPIRou bandpass. Low-impact-parameter transits would lead to first an apparent increase of temperature (planet occulting the limb) and a decrease of the mean temperature when the planet occults regions close to the disk center. Grazing transits lead to a positive excursion only. [3] for the nominal impact parameter ($b=0.69$) and a 2.8\% spot filling factor \citep{sing_hubble_2011}, we show the predicted \dtemp signal through transit [3] as well as the corresponding broadband lightcurve [4].}
    \label{fig:toymodel_predictions}
\end{figure*}

\subsubsection{HD\,189733 transit \dtemp time series \label{sect:hd189733}}
\begin{figure*}[!htbp]
    \includegraphics[width=\linewidth]{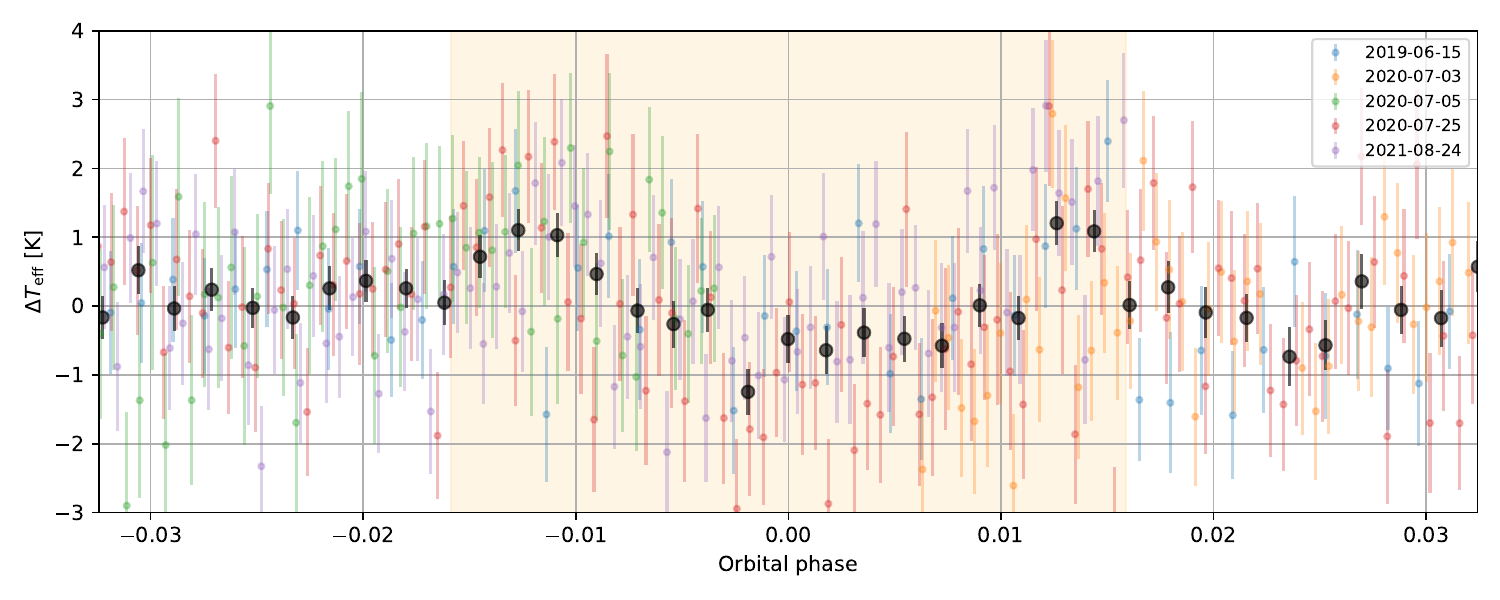}
    \caption{\dtemp time series of four HD189733\,b transit observed with SPIRou and phase-binned values. One notes the $\sim$1\,K increase in the apparent temperature of the host star close to ingress and egress as cooler regions of the stellar disk are masked by the planet. The beige domain corresponds to the window between 1$^{\rm st}$ and $4^{\rm th}$ contact for an impact parameter of 0.69.}
    \label{fig:renard}
\end{figure*}

Figure~\ref{fig:renard} overplots 5 transits of HD\,189733\,b in the temperature space. Each time series is subtracted from a baseline between 1.8 and 1.0 half-transit durations on either side of the event. Phase-binned points are overplotted and one sees a temperature excess just after ingress and just before egress. One sees an increase in \dtemp of $\sim$1\,K just after ingress and just before degree, with a $\sim0.5$\,K decrease below the baseline \dtemp value at mid-transit. Qualitatively, this is understood as an increase in the disk-averaged temperature when the limb is occulted and a decrease in disk-averaged temperature, at mid-transit, when central regions of the disk are occulted.  This \dtemp variation through transit is analogous to the RM, but in temperature rather than velocity space. 

\begin{figure}[!htbp]
    \includegraphics[width=\linewidth]{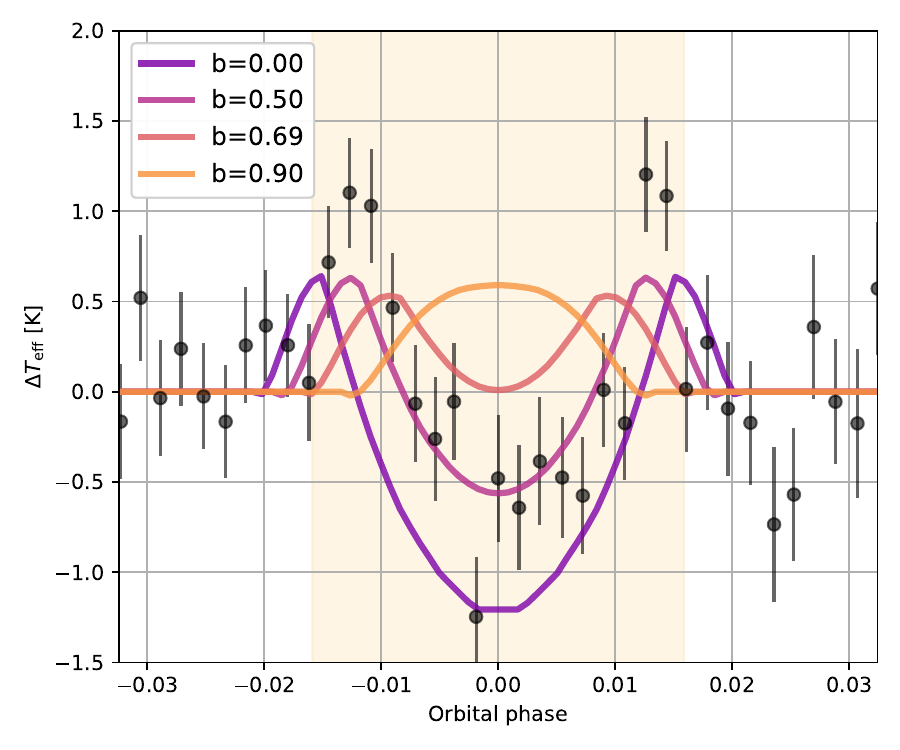}
    \caption{Binned \dtemp measurements through HD\,189733\,b transits compared with a change in the impact parameter. The $\sim$0.5\,K dip in \dtemp compared to out-of-transit values as well as the maximum in \dtemp at a position of $\pm$0.8 stellar radii would suggest an impact parameter of $b\sim$0.5, which is significantly smaller than the literature value ($b=0.69$). Here, we assume that the stellar photosphere is uniform (i.e., no spots, no plages).}
    \label{fig:renard_b}
\end{figure}

\begin{figure}[!htbp]
    \includegraphics[width=\linewidth]{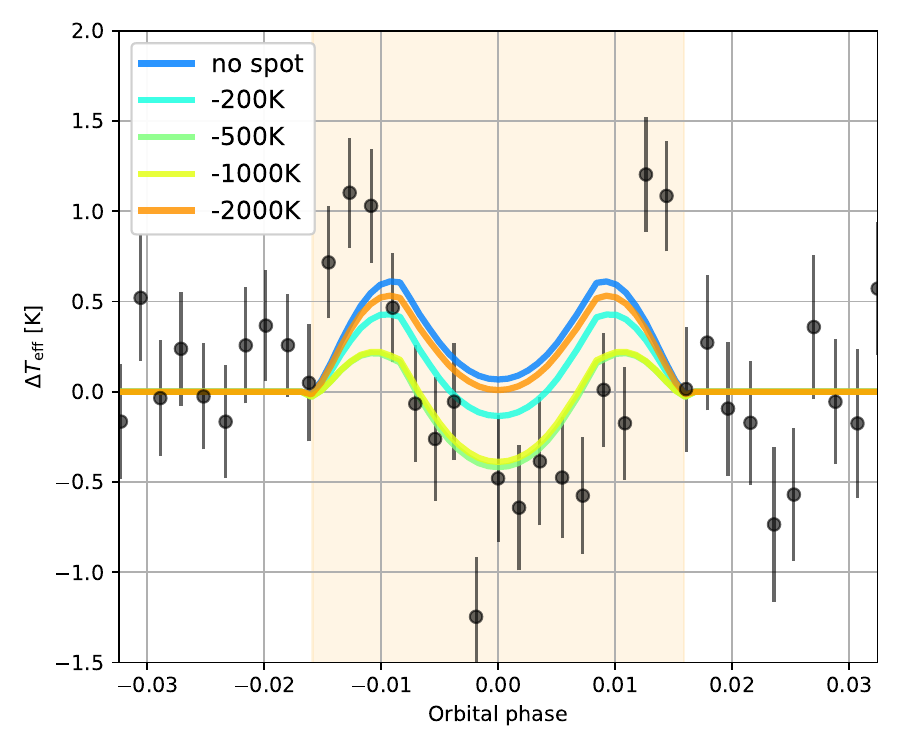}
    \caption{Binned \dtemp measurements through HD\,189733\,b transit compared with a change in the temperature contrast of spots covering 2.8\% of the photosphere. We assume that spots are outside of the transit chord (as in Figure~\ref{fig:toymodel}). The dip below the baseline value at mid-transit suggests a starspot contrast of $\sim1000$\,K, but this model strongly underpredicts {the rise} at the start and end of the transit. We note that having very cold spots ($-2000$\,K) or no spots leads to a very similar signature; very cold spots contribute little to the overall spectrum and have no effects on the \dtemp signal.}
    \label{fig:renard_spot_teff}
\end{figure}

While this qualitative behaviour is reproduced by the model described in Section~\ref{sec:model_dtemp}, no simple model reproduced quantitatively the observed \dtemp signal. Figure~\ref{fig:renard_b} illustrates the binned \dtemp signal in comparison with the predicted signature varying the impact parameter of HD\,189733\,b. The overall behaviour is best reproduced with an impact parameter of $b\sim0.5$, which leads to a lower mid-transit minimum than the literature impact parameter. Considering that the impact parameter is very well constrained by space-based photometry and accurate pRV work (e.g., {\cite{cristo_espresso_2024}}), we do not claim that literature values of the impact parameter are erroneous, but other effects must be at play. Figure~\ref{fig:renard_spot_teff} illustrates the impact of unocculted spots with varying temperatures. The central `dip' in the \dtemp signature is well reproduced with a contrast of $\sim$1000\,K, which is also consistent with the temperature contrast derived by \citet{sing_hubble_2011}. While this unocculted spot model matches the central part of the transit, it fails to reproduce the amplitude of the \dtemp increase just after ingress and just before egress. No unocculted spot models reproduce the $\sim$1\,K  increase in \dtemp close to ingress and egress and their maxima happens too close to mid-transit compared to the SPIRou time series. No simple combination of parameters (spot contrast and distribution) that assumed a transit geometry consistent with literature values has been found to fit the observed \dtemp sequence. Understanding the underlying reasons for this mismatch is beyond the scope of the current work. The HD\,189733 system has been observed with several high-resolution spectrographs (e.g., SOPHIE, HARPS, HIRES, ESPRESSO; \cite{lanza_deriving_2011,cauley_decade_2017, cristo_espresso_2024}). The joint analysis of the \dtemp effect in these datasets, along with the SPIRou observations presented here, may shed light on the origin of this discrepancy between simplistic models and the observed transit \dtemp effect. It remains to be seen if a more complex geometry (e.g., combining spots and plages) or if inconsistencies in stellar atmosphere models should be invoked.

Spot and plage-crossing events should also manifest themselves as transit-induced temperature changes. These are readily detectable in photometric and spectro-photometric data but may be harder to identify in ground-based, high-resolution spectroscopic transit observations. While no such event is seen in the dataset in hand, star spot crossing events should manifest themselves as a positive \dtemp excursion. The peculiar \hbox{TOI-3884} system with a persistent star spot transited by a super-Neptune \citep{almenara_toi-3884_2022, libby-roberts_-depth_2023} would be a particularly compelling case study. Obtaining simultaneous photometric and \dtemp measurements would provide additional constraints regarding spot filling ratio and temperature contrast. Overall, the \dtemp through transit provides a new tool to quantify the transit light source effect \citep{rackham_transit_2018} emerging as a significant issue in JWST transit spectroscopy (e.g., \cite{lim_atmospheric_2023}).

\section{Models versus empirical \dtemp spectra\label{sect:models_vs_templates}}
The usage of empirical temperature sequences, as described in Section~\ref{sect:method}, implies that one must assemble a sizable library of archival spectra of stars at known temperatures. Ideally, a \dtemp spectrum could be derived with models and applied to the time series, alleviating this requirement. We explored this approach with spectra using PHOENIX \citep{husser_new_2013} spectral library.  
\citet{shahaf_linearized_2023} also considered the spectral variation of spectra with wavelength. They explore the representation of radial velocity shifts in the context of Taylor expansions relative to various quantities, including velocity (the RV shift) and temperature (equivalent to $\partial A_0 /\partial T$ here).

\begin{figure*}[!htbp]
    \includegraphics[width=\linewidth]{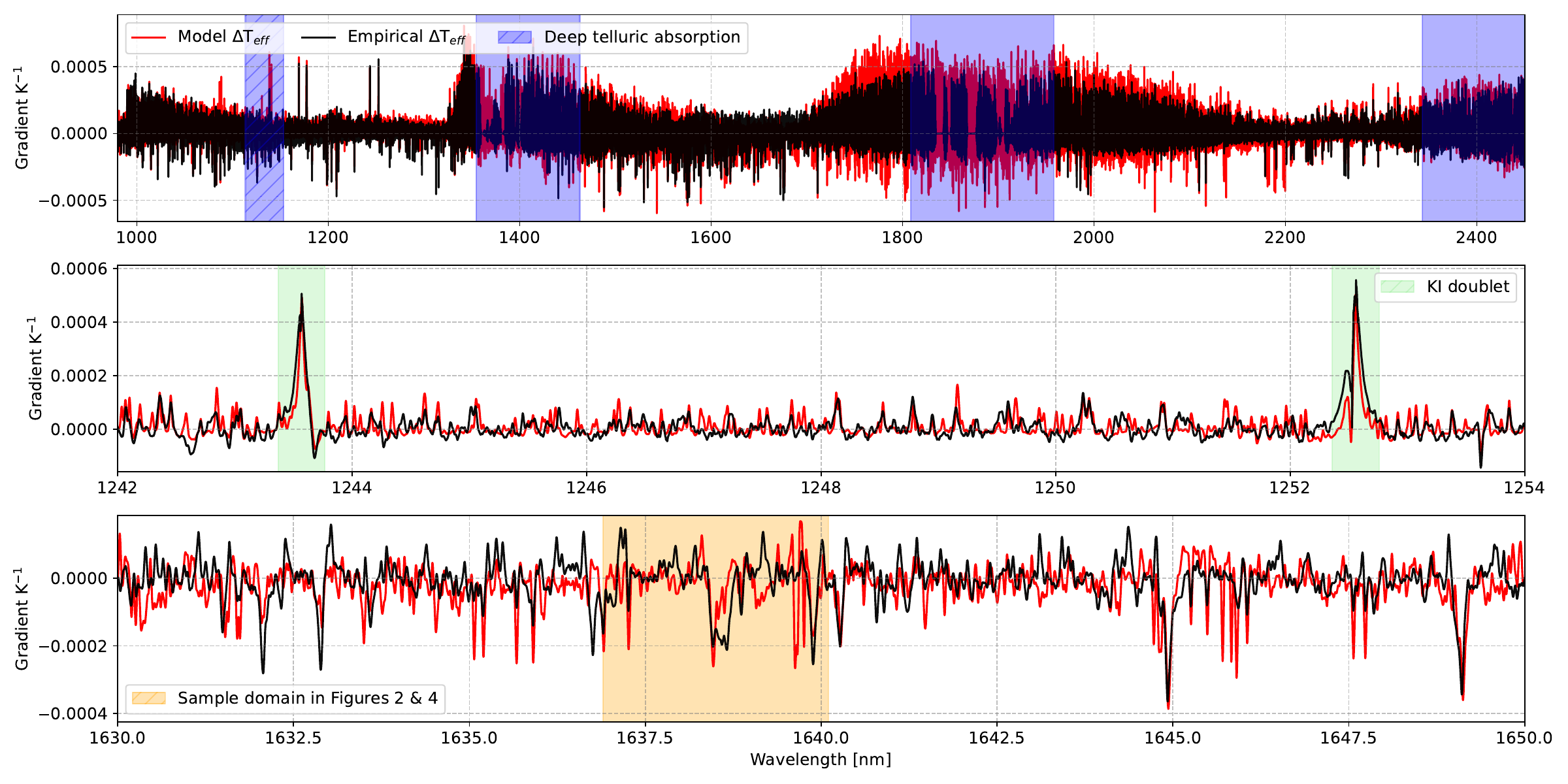}

    \caption{ Empirical and model temperature gradient spectra for the whole SPIRou domain (top), centred on the $J$-band KI doublet (middle) and a sample domain within $H$ band for T$_{\rm eff}$=3500\,K. The overall behaviour of the temperature gradient is well reproduced and most of the deep lines are reproduced by models. One notes that numerous small features, particularly in the $H$ band, are not accurately predicted by models.}
    \label{fig:dtemp_spectra_model_obs}
\end{figure*}

A \dtemp spectrum was obtained by applying the high-pass filter to two spectra at increasing temperatures and applying the same method as for spectral templates. Figure~\ref{fig:dtemp_spectra_model_obs} shows the temperature gradient at 3500\,K for both theoretical and empirical spectra. The overall behaviour of the temperature gradient is reproduced, as well as strong features such as the KI doublet. Numerous small features are not reproduced correctly, which is consistent with the fact that the details of M dwarf spectra are, generally, poorly reproduced at high spectral resolution, and one-to-one identification of lines can be challenging \citep{jahandar_comprehensive_2023, artigau_optical_2018}.

 \begin{figure*}[!htbp]
    \includegraphics[width=\linewidth]{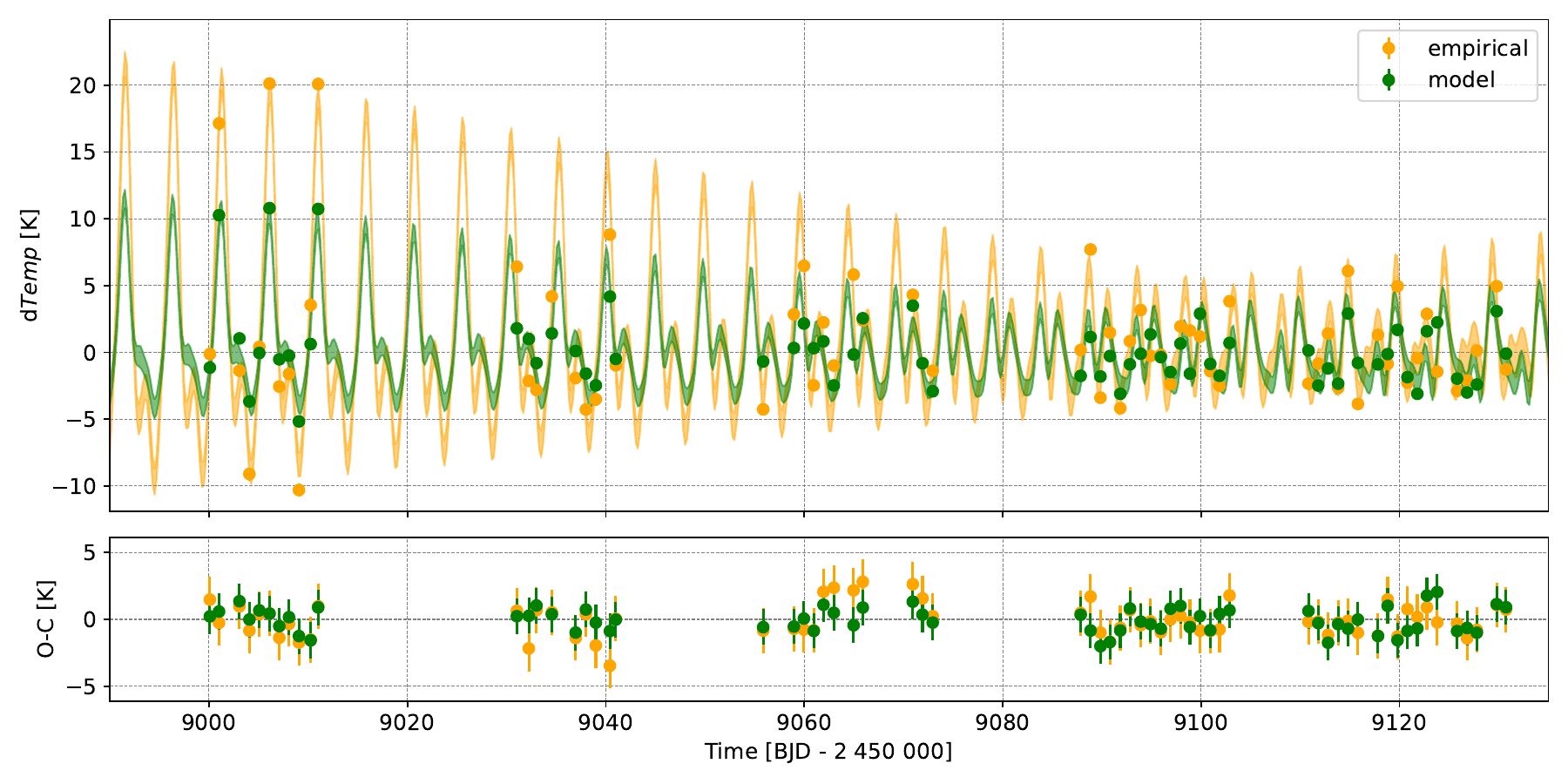}
    \caption{\dtemp time series with empirical and model temperature gradient spectra for the same timespan as in Figure~\ref{fig:gp_aumic}. The signature of the activity is recovered but with an amplitude reduced by a factor $\sim$2 while the periodicity and correlation length of the GP fits are consistent.}
    \label{fig:aumic_model_dtemp}
\end{figure*}

Deriving a \dtemp time series with a template that only partially correlates with actual temperature gradients will lead to a lower apparent temperature change as well as a loss in the significance of the detection. We verified this by using the theoretical temperature gradient shown in Figure~\ref{fig:dtemp_spectra_model_obs} on the AU Mic time series, with the results shown in Figure~\ref{fig:aumic_model_dtemp}. While the overall behaviour is preserved and one can retrieve the rotation period with the same accuracy as with the empirical temperature gradient. The GP fit on the time series derived from model gradients gives a rotation period of \postLINAUMicModellnP\,days, a scale length of \postLINAUMicModellnl\,days and shape factor $\Gamma=$\postLINAUMicModellngamma, all being consistent with values reported in Table~\ref{tbl:aumic}. The derived amplitude $\sigma_{\rm GP} =$\postLINAUMicModellnA\,K is a factor $\sim$2 smaller. This difference in amplitude arises from the fact that the model \dtemp spectrum only partially traces the empirical effect of a temperature change.

\begin{figure}[!htbp]
    \includegraphics[width=\linewidth]{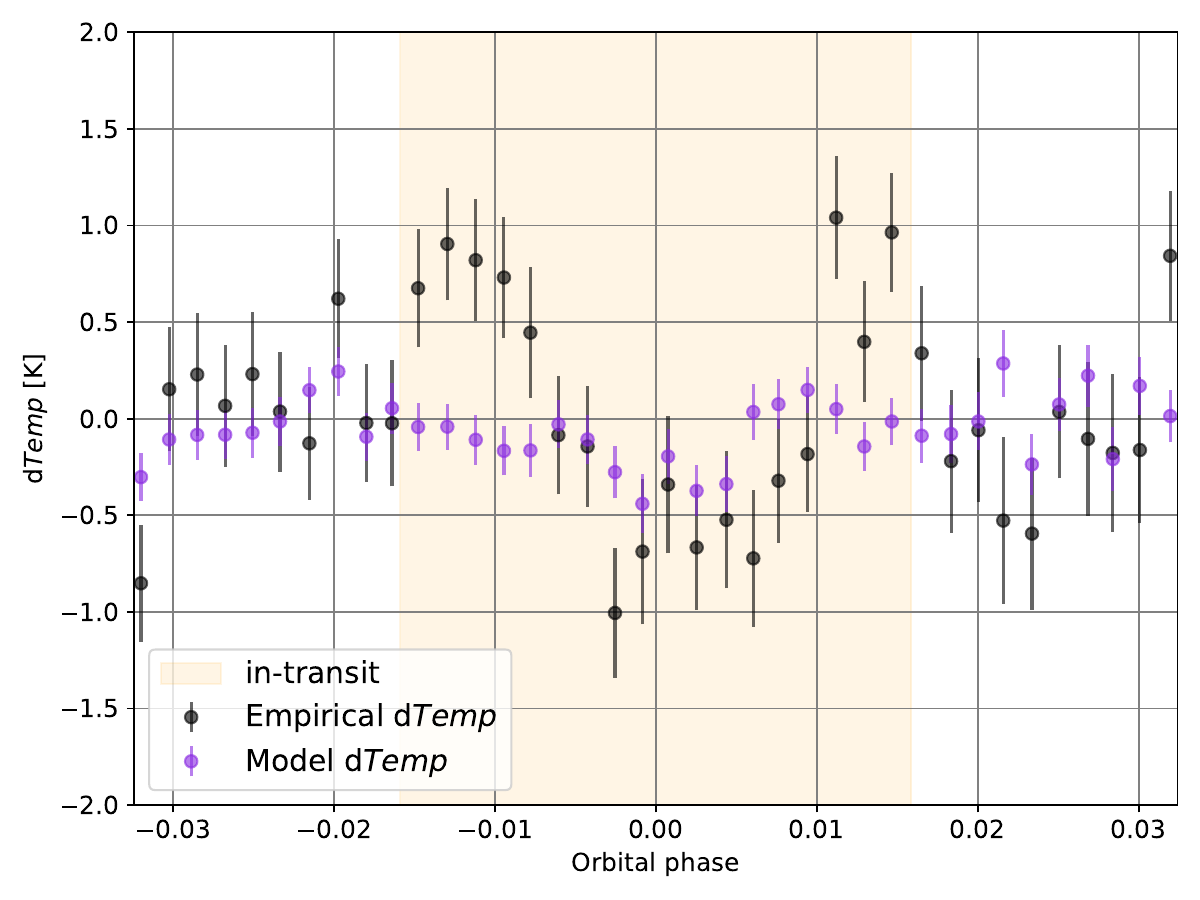}
    
    \caption{{Same as for Figure~\ref{fig:renard}, but comparing the effect of using empirical and model-based gradient spectra. The overprediction of temperature gradients in models leads to artificially small uncertainties, but the signature of the effect described in Section~\ref{sect:hd189733} disappears.} }
    \label{fig:hd189733_model_dtemp}
\end{figure}

We determined the \dtemp with model spectra for HD189733; Figure~\ref{fig:hd189733_model_dtemp} shows the \dtemp signal with both empirical and model temperature gradient spectral. The effect described in Section~\ref{sect:hd189733} disappears completely and one gets a much flatter time series. The median uncertainties are reduced from \mederrHDdtemp\,K to \mederrHDmodel\,K as models overpredict temperature gradients at 5000\,K, but this comes with a true loss in the accuracy of the measurements. Overall, empirical gradients are strongly preferred for this type of analysis and \dtemp values retrieved in this framework with model-based gradients should be viewed, at best, with caution.

\section{Conclusions and future prospects\label{sect:conclusions}}
We highlighted with sample datasets the interest of a new stellar activity indicator, the \dtemp value, in time series of high-resolution spectroscopy, in particular in the context of pRV exoplanet searches. The indicators provide a measurement of {\it differential} temperature of the star in a time series and, for bright targets, can reach uncertainties well below 1\,K per measurement. The activity indicator relies on a library of spectral templates of known temperature, but this library only needs to be determined once for a given combination of spectral type and wavelength domain (See Appendix~\ref{data_availability} for access to the templates). Therefore, new instruments do not have to re-observe a large set of templates to measure this indicator.

The framework opens a new window in stellar astrophysics, allowing for very subtle changes in {disk-averaged temperatures} to be measured, at the sub-K level. If one expresses this as the equivalent of a photometric change, a 1\,K temperature change on a 4000\,K black body (roughly the temperature of an M0) corresponds to a photometric variation of 1\,mmag (assuming a $T_{\rm eff}^4$ dependency of flux). Obtaining sub-mmag photometry simultaneously with pRV observations, over baselines of months to years is virtually impossible from the ground; sub-mmag ground-based photometry from the ground has only been demonstrated within nights (e.g., \citealt{stefansson_toward_2017}). With \dtemp measurements, the time sampling is contemporaneous with pRV measurement as both are drawn from the same dataset. Future work will explore the equivalent of the $f \cdot f'$ \citep{aigrain_simple_2012} framework but using temperature rather than the flux to correct for activity jitter in pRV sequences.

It may be non-trivial to determine the underlying physical processes traced by a given measured  \dtemp signal. In the case of AU\,Mic, the notable correspondence between small scale \magb and \dtemp points toward the magnetic origin of the temperature variations, but these may be purely geometric in the case of HD\,189733. Stellar pulsations also probably lead to detectable \dtemp for G and K dwarfs, a case that has not been investigated here. One more subtle effect may be unrelated physical causes, such as variations in the Zeeman splitting from the magnetic field, that lead to a modification of line profiles with a non-zero cross-term with the temperature derivative spectrum. Changes in chemical abundances, leading to a change in line depth, may also lead to an erroneous \dtemp interpretation. Another plausible mechanism for \dtemp signatures is through planet-star interaction that has yet to be explored. Close-in planets (e.g., \citealt{strugarek_chasing_2019, strugarek_diversity_2014}) interact with the large-scale magnetic field of the host star, as the planet rotates inside the magnetosphere. This second-order effect in $\Delta T_{\rm eff}$ may be detectable as a modulation at the orbital period of the planet. Detecting a periodicity peak in \dtemp at the orbital period of a tentative radial-velocity detection, or its higher harmonics should therefore be seen with caution. It may point to a false positive of stellar activity masquerading as a planetary signal, but it may also be indicative of planet-star interaction.

The framework presented here highlights the usage of the \dtemp framework in the context of high-resolution spectroscopy at optical and infrared wavelengths. Conceptually, the framework should work at any spectral resolution, provided that the signal-to-noise ratio is sufficiently high. Transit-spectroscopy time series with JWST are particularly appealing for a low-resolution application of the \dtemp framework. As pointed out earlier, the transit light source effect is a significant limitation to transit spectroscopy and must be accounted for in the interpretation of transit spectroscopy data, in particular at the bluer end of the JWST domain. \cite{lim_atmospheric_2023, doyon_temperate_2024}{ and \cite{cadieux_transmission_2024}} illustrate that for mid-to-late M dwarfs, the domain shortward of $\sim$1.0\,$\mu$m becomes dominated by these effects. Out-of-transit \dtemp effects in JWST data could point to rotation-induced temperature modulation and, when combined with raw photometry, inform on the filling factor of active regions. In-transit \dtemp effects, analogous to that of HD\,189733\,b shown in Figure~\ref{fig:renard}, would inform on light source effects and provide a strong prior for the models used in the interpretation of transit spectroscopy data. One limiting factor in constructing a \dtemp framework for JWST transit data is the lack of a large spectral library, uniformly stepped in effective stellar temperatures, obtained with each observation mode. One can expect such a library to become gradually available as more JWST transit data becomes public in the next few years.

\acknowledgments

Based on observations obtained at the Canada-France-Hawai\okina i  Telescope which is operated from the summit of Maunakea by the National Research Council of Canada, the Institut National des Sciences de l'Univers of the Centre National de la Recherche Scientifique of France, and the University of Hawai\okina i . The observations at the Canada-France-Hawai\okina i  Telescope were performed with care and respect from the summit of Maunakea which is a significant cultural and historic site. This work is partly supported by the Natural Science and Engineering Research Council of Canada, the Fonds Qu\'eb\'ecois de Recherche  (Nature et Technologie) the Trottier Family Foundation through the Institute for Research on Exoplanets.

EA, CC, NJC, RD, MC \& LD acknowledge the financial support of the FRQ-NT through the {\it Centre de recherche en astrophysique du Québec} as well as the support from the Trottier Family Foundation and the Trottier Institute for Research on Exoplanets 

JFD, CM and PC acknowledge funding from the European Research Council (ERC) under the H2020 research \& innovation programme (grant agreement \#740651 New- Worlds).

R.A. acknowledges the Swiss National Science Foundation (SNSF) support under the Post-Doc Mobility grant P500PT\_222212 and the support of the Institut Trottier de Recherche sur les Exoplanètes (iREx).

\clearpage 
This research made use of the following software tools:
\begin{itemize}
    \item \texttt{Astropy}; a community-developed core Python package for Astronomy. \citet{the_astropy_collaboration_astropy_2013,the_astropy_collaboration_astropy_2018}.

    \item \texttt{george}; \citet{ambikasaran_fast_2016}.
    \item \texttt{IPython}; \citet{perez_ipython_2007}.
    \item \texttt{matplotlib}, a Python library for publication quality graphics; \citet{hunter_matplotlib_2007}. 
    \item \texttt{NumPy}; \citet{harris_array_2020}.
    \item \texttt{radvel}; \citet{fulton_radvel_2018}.
    \item \texttt{SciPy}; \citet{scipy_10_contributors_scipy_2020}.
    \item \texttt{tqdm}; \citet{da_costa-luis_tqdm_2021}.
    \item During the analysis work, EA and NC made extensive use of the Dace platform \citep{buchschacher_data_2015}.
\end{itemize}

\facilities{CFHT(SPIRou), ESO La Silla (HARPS), ESO Science Archive.}

\appendix
\counterwithin{table}{section}
\counterwithin{figure}{section}

\section{Data availability\label{data_availability}}

The template derivatives used here are provided for the HARPS and SPIRou domains in the LBL webpage\footnote{\url{https://lbl.exoplanets.ca}}. We plan to upgrade these templates with full coverage of the 400\,nm to 2450\,nm domain with archival data from instruments covering the far-red domain. The \dtemp indicator is computed for temperatures ranging from 3000\,K to 6000\,K in steps of 500\,K and for rotational broadenings ranging from 0 to 19\,km/s in steps of 1\,km/s. The LBL framework now accepts data from several instruments (SPIRou, NIRPS, SOPHIE, HARPS, HARPS-N, CARMENES-VIS, ESPRESSO, MAROON-X), and other instruments may be added upon reasonable request and pending resource availability.

\begin{table}[!htbp]
\caption{First entries in the full table, available in electronic form, of the templates used for the current work. A total of \nstarsharps\ have been used with HARPS and \nstarsspirou\ with SPIRou, with \nstarsoverlap\ in common between the two instruments.}\label{tbl:models}
\begin{tabular}{lcrccc}
\hline
{SIMBAD$^a$} identifier & Instrument(s) & $T_{\rm eff}$ & Fe/H & $\log g$ & $V_{\rm sys}$ \\
&&[K]&[dex]&[dex]&m/s\\
\hline
G 157-77 & SPIRou & 2863 & -0.47 & 5.10 & -41066 \\
PM J22114+4059 & SPIRou & 2919 & -0.04 & 5.07 & -16835 \\
G 141-36 & SPIRou & 2929 & -0.21 & 5.11 & -34544 \\
LP  791-18 & SPIRou & 2930 & -0.23 & 5.11 & 14545 \\
L  788-37 & HARPS & 3061 & -0.13 & 5.05 & -24292 \\
G 112-50 & SPIRou & 3075 & -0.05 & 5.02 & 38704 \\
G  36-24 & SPIRou & 3089 & -0.31 & 4.86 & -6489 \\
Ross  128 & HARPS, SPIRou & 3093 & -0.20 & 5.06 & -31111 \\
TOI-1452 & SPIRou & 3170 & -0.03 & 4.98 & -34471 \\
BD-15  6290 & HARPS, SPIRou & 3185 & -0.00 & 4.78 & -1490 \\
\hline
\footnotesize{$^a$ \cite{wenger_simbad_2000}}\\

\end{tabular}
\end{table}

\bibliography{references.bib}{}
\bibliographystyle{aasjournal.bst}
\end{document}